\documentclass[iop,numberedappendix,appendixfloats,twocolappendix]{emulateapj}

\usepackage{subfigure}
\usepackage{graphicx}
\usepackage[usenames]{color}

\received{}
\revised{}
\accepted{}

\shortauthors{Boyer et al.}
\shorttitle{``DUSTINGS Overview''}

\begin{document}

\title{An Infrared Census of DUST in Nearby Galaxies with Spitzer (DUSTiNGS), I. Overview}

\author{Martha~L.~Boyer\altaffilmark{1,2},
  Kristen~B.~W. McQuinn\altaffilmark{3}, Pauline Barmby\altaffilmark{4},
  Alceste~Z. Bonanos\altaffilmark{5}, Robert D. Gehrz\altaffilmark{3},
  Karl~D.~Gordon\altaffilmark{6}, M.~A.~T. Groenewegen\altaffilmark{7},
  Eric Lagadec\altaffilmark{8}, Daniel Lennon\altaffilmark{9}, Massimo
  Marengo\altaffilmark{10}, Margaret~Meixner\altaffilmark{6}, Evan
  Skillman\altaffilmark{3}, G.~C.~Sloan\altaffilmark{11},
  George~Sonneborn\altaffilmark{1},
  Jacco~Th.~van~Loon\altaffilmark{12},
  Albert~Zijlstra\altaffilmark{13}} 
\altaffiltext{1}{Observational
  Cosmology Lab, Code 665, NASA Goddard Space Flight Center,
  Greenbelt, MD 20771 USA; martha.boyer@nasa.gov} 
\altaffiltext{2}{Oak
  Ridge Associated Universities (ORAU), Oak Ridge, TN 37831 USA}
\altaffiltext{3}{Minnesota Institute for Astrophysics, School of Physics and Astronomy, 116 Church Street SE, University of
  Minnesota, Minneapolis, MN 55455 USA}  
\altaffiltext{4}{Department of Physics \& Astronomy, University of Western Ontario, London, ON, N6A 3K7, Canada}
\altaffiltext{5}{IAASARS, National Observatory of Athens, GR-15236 Penteli, Greece}
\altaffiltext{6}{STScI, 3700 San Martin Drive, Baltimore, MD 21218
  USA}
\altaffiltext{7}{Royal Observatory of Belgium, Ringlaan 3, B-1180 Brussels, Belgium}
\altaffiltext{8}{Laboratoire Lagrange, UMR7293, Univ. Nice Sophia-Antipolis, CNRS, Observatoire de la C\^{o}te d'Azur, 06300 Nice, France}
\altaffiltext{9}{ESA - European Space Astronomy Centre, Apdo. de Correo 78, 28691 Villanueva de la Ca\~{n}ada, Madrid, Spain}
\altaffiltext{10}{Department of Physics and Astronomy, Iowa State University, Ames, IA 50011, USA}
\altaffiltext{11}{Astronomy Department, Cornell University, Ithaca, NY 14853-6801, USA}
\altaffiltext{12}{Astrophysics Group, Lennard-Jones Laboratories,
  Keele University, Staffordshire ST5 5BG, UK}
\altaffiltext{13}{Jodrell Bank Centre for Astrophysics, Alan Turing Building, University of Manchester, M13 9PL, UK}

\begin{abstract}
Nearby resolved dwarf galaxies provide excellent opportunities for
studying the dust-producing late stages of stellar evolution over a
wide range of metallicity ($-2.7 \lesssim [{\rm Fe/H}] \lesssim
-1.0$). Here, we describe DUSTiNGS (DUST in Nearby Galaxies with {\it
Spitzer}): a 3.6 and 4.5~\micron\ post-cryogen {\it Spitzer Space
Telescope} imaging survey of 50 dwarf galaxies within 1.5~Mpc that is
designed to identify dust-producing Asymptotic Giant Branch (AGB)
stars and massive stars. The survey includes 37 dwarf spheroidal, 8
dwarf irregular, and 5 transition-type galaxies. This near-complete
sample allows for the building of statistics on these rare phases of
stellar evolution over the full metallicity range. The photometry is
$>$75\% complete at the tip of the Red Giant Branch for all targeted
galaxies, with the exception of the crowded inner regions of IC\,10,
NGC\,185, and NGC\,147. This photometric depth ensures that the
majority of the dust-producing stars, including the thermally-pulsing
AGB stars, are detected in each galaxy. The images map each galaxy to
at least twice the half-light radius to ensure that the entire evolved
star population is included and to facilitate the statistical
subtraction of background and foreground contamination, which is
severe at these wavelengths. In this overview, we
describe the survey, the data products, and preliminary results. We
show evidence for the presence of dust-producing AGB stars in 8 of the
targeted galaxies, with metallicities as low as ${\rm [Fe/H]} =
-1.9$, suggesting that dust production occurs even at low
metallicity.

\end{abstract}


\vfill\eject
\section{INTRODUCTION}
\label{sec:intro}

\subsection{Dust Production by Evolved Stars}

Intermediate-mass ($1\,M_\odot \lesssim M\lesssim 8\,M_\odot$) and
massive ($\gtrsim$8~$M_\odot$) evolved stars are drivers of galaxy
chemical enrichment and evolution via the return of significant
amounts of gas and dust to the interstellar medium (ISM). This stellar
mass loss also drives the subsequent evolution of the stars
themselves. However, post-main sequence stellar evolution is poorly
understood, especially in the short-lived dust-producing
phases. And it is unclear how the galactic environment
(especially metallicity) affects stellar dust production and
evolution. DUST in Nearby Galaxies with {\it Spitzer} (DUSTiNGS) is an
infrared (IR) survey of 50 dwarf galaxies in and around the Local
Group designed to detect evolved stars in the dust-producing phase.

Massive dusty evolved stars such as luminous blue variables,
Wolf-Rayet stars, red supergiants, and supergiant B[e] stars are
prolific dust producers \citep{Smith2014,Bonanos+2010, Kastner+2006,
  Voors+2000, Smith+2003}, though it is uncertain how much, if any,
dust will survive the subsequent supernova (SN) explosion. The role of
episodic mass loss, which is often accompanied by dust production, in
the evolution of massive stars remains an open question.  The inferred
presence of pre-existing circumstellar material around several
core-collapse SNe \citep{Smith+2007} suggests that mass loss plays an
important part in stellar evolution. The DUSTiNGS survey includes a
large sample of nearby dwarf galaxies to increase
the known sample of these short-lived stars over a wide
range of stellar masses and metallicities.

Intermediate-mass Asymptotic Giant Branch (AGB) stars condense dust
from material formed in situ and may be a major source of interstellar
dust \citep{Gehrz89} as inferred, for example, by the AGB origin of a
large fraction of presolar grains found in meteorites
\citep[e.g.,][]{Gail+2009}.  Several works have shown that a small
population of very dusty AGB stars dominate the AGB dust production in
the Magellanic Clouds at a given time \citep{Srinivasan+2009,
  Boyer+2012, Riebel+2012, Zhukovska+2013, Schneider+2014}. These
stars (sometimes called ``extreme'' AGB stars --- or x-AGB stars) are
optically obscured, and are generally selected via their red colors
($[3.6]-[8] > 3$~mag; see Section~\ref{sec:xagb_class}).  They
comprise $\lesssim$5\% of the AGB population, but produce more than
$3/4$ of the AGB dust. Through spectral energy distribution modeling,
\citet{Riebel+2012} find that most of these stars in the Large
Magellanic Cloud are carbon rich. Carbon stars form easily in
metal-poor environments because of a low initial oxygen abundance in the
circumstellar envelope and hence more free carbon after the formation
of C/O \citep[e.g.,][]{GroenewegendeJong1993}. It follows that carbon
stars may also dominate the dust production in the more metal-poor
dwarf galaxies.

Nevertheless, it is still unclear how much AGB dust survives the harsh
environment of the ISM produced by SN shocks
\citep[e.g.,][]{JonesNuth2011}. Recent {\it Spitzer} observations of
the Small and Large Magellanic Clouds (SMC/LMC) by the Surveying the
Agents of Galaxy Evolution program
\citep[SAGE;][]{Meixner+06,Gordon+11} produced a complete census
of AGB stars in those galaxies. Estimates of the total dust input
compared with other known dust sources (i.e., supernovae ejecta)
indicate that AGB stars may be the dominant source of stellar-produced
dust grains \citep{Matsuura+09, Boyer+2012, Riebel+2012,
  Zhukovska+2013, Schneider+2014}. These works also concluded that,
despite their efficient dust production, AGB stars can account for
only a fraction of the ISM dust mass in the SMC and LMC.  However, a
revised measurement of the SMC and LMC ISM dust masses using {\it
  Herschel Space Observatory} data is significantly smaller than
previously estimated with {\it Spitzer} data \citep{Gordon+2014},
indicating that AGB stars may in fact be a dominant dust source in
these galaxies.

\subsection{The Metallicity Dependence of Dust Production}
\label{sec:agb}

For more metal-poor populations, the metallicity dependence of dust
production by AGB stars remains unclear.  Some AGB stellar evolution
models suggest that dust production easily occurs at very low
metallicity because carbon stars create carbon in situ
\citep[e.g.,][]{Karakas+2007,Mattsson+2008,Wachter+2008}. Other models
suggest that at very low metallicity (${\rm [Fe/H]} \lesssim -2$), AGB
stars contribute little dust and thus provide a negligible
contribution to the total dust budget of high-redshift galaxies
(L. Mattsson, in preparation).

The effect of the metallicity on dust production likely
  differs for oxygen-rich and carbon-rich AGB stars because carbon stars
  create their own carbon. Photometric surveys of metal-poor globular
clusters show modest dust-production by low-mass oxygen-rich AGB stars
\citep{Boyer+09a,McDonald+2011a,McDonald+2011b} down to ${\rm [Fe/H]}
\approx -1.7$. Infrared spectroscopy of O-rich AGB stars in the
Magellanic Clouds and globular clusters reveals trends consistent with
reduced dust production at lower metallicities, as expected due to
reduced amounts of the oxygen needed to make silicate dust, but these
studies are not conclusive
\citep[][]{Sloan+08,Groenewegen+09,Sloan+2010}.

In carbon stars, some works suggest there is no metallicity dependence
on dust formation \citep[e.g.,][]{Groenewegen+07}, while some do find
hints of such a dependence at ${\rm [Fe/H]} \lesssim -1$
\citep{vanLoon+08b,Sloan+2012}. However, this latter group includes only
two C stars in the Sculptor dwarf and three in Leo\,I
\citep[${\rm [Fe/H]} = -1.68$ and $-1.43$,
  respectively;][]{McConnachie+2012}. Larger samples at low
metallicities are clearly needed. 

\begin{figure}
  \includegraphics[width=\columnwidth]{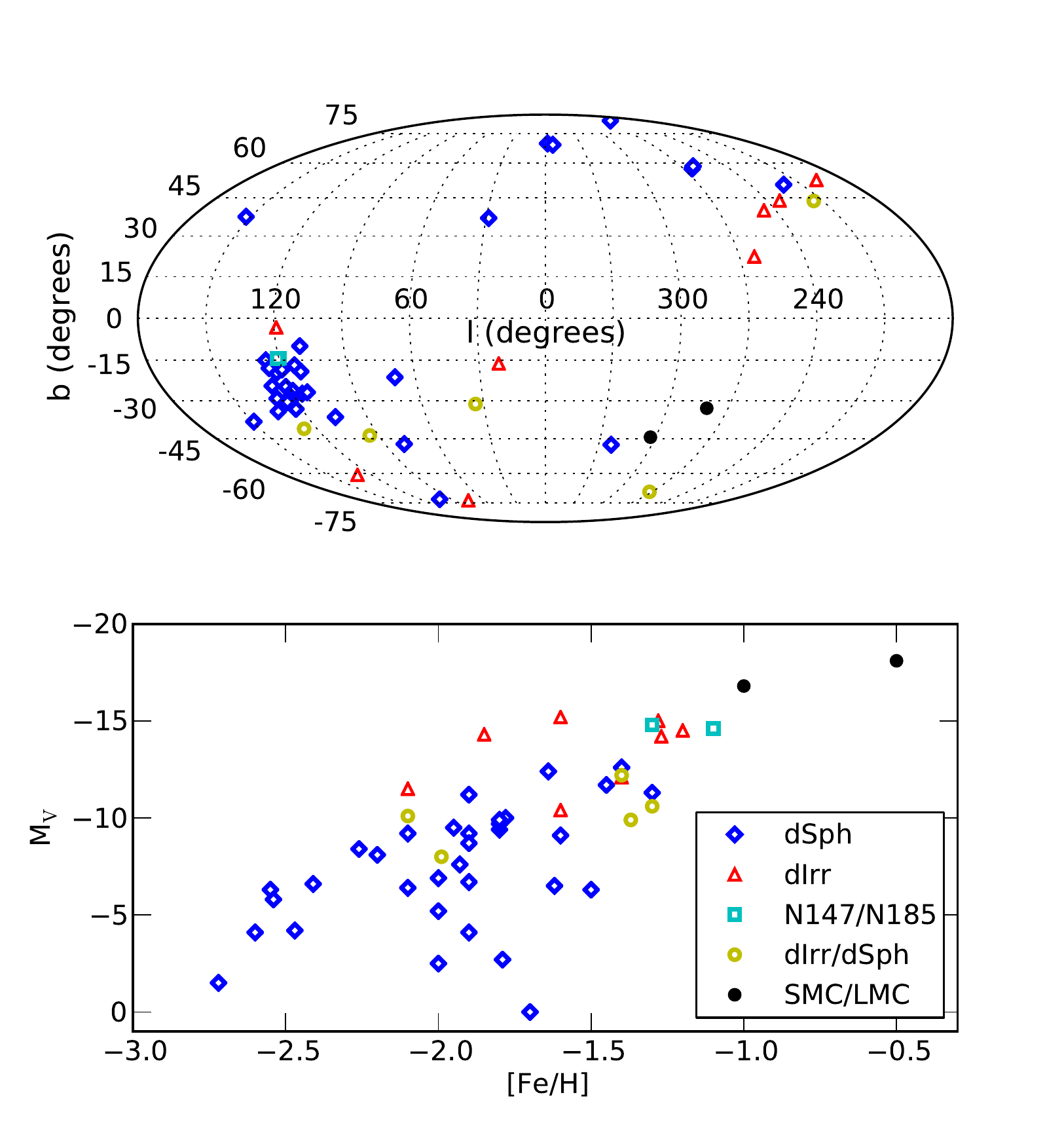}
\caption{DUSTiNGS targets with properties from
  Table~\ref{tab:targets}. {\it Upper panel:} Distribution of target
  galaxies in Galactic coordinates. Note there are few targets
  near the Galactic Plane/Bulge, limiting the effects of foreground
  extinction and contamination (Sections~\ref{sec:ext} and
  \ref{sec:cmd}). The cluster of dSph galaxies near $l=120\degr$ and
  $b=20\degr$ is the Andromeda group. {\it Lower panel:} Distribution
  of target galaxies in absolute $V$-band magnitude ($M_{\rm V}$) and
  metallicity ([Fe/H]). The Small and Large Magellanic Clouds
  (SMC/LMC) are shown for comparison. \label{fig:targs}}
\end{figure}

The DUSTiNGS survey aims to build statistics of the short-lived
dust-producing phase at low metallicity for constraining stellar
evolution and dust production models. Here, we present an overview of
the survey, which greatly extends the baseline in age and metallicity
over previous observations (Table~\ref{tab:targets},
Fig.~\ref{fig:targs}), and provides a near-complete census of galaxies
within 1.5~Mpc at 3.6 and 4.5~\micron.  The purpose of this overview
is to describe the DUSTiNGS targets (Section~\ref{sec:targets}), the
observations and survey design (Section~\ref{sec:obs}), and the data
products (Sections~\ref{sec:phot} and \ref{sec:cat}). We also estimate
the AGB population size (Section~\ref{sec:pops}).  Forthcoming papers
will describe additional scientific results in detail; in
\citet[][hereafter Paper\,II]{Boyer+2014b}, we identify individual
x-AGB star candidates via their pulsation.

\begin{deluxetable*}{rlllllllc}
\tablewidth{0pc}
\tabletypesize{\normalsize}
\tablecolumns{9}
\tablecaption{Adopted Target Parameters\label{tab:targets}}

\tablehead{
\colhead{Galaxy}&
\colhead{R.A.}&
\colhead{Dec}&
\colhead{$(m-M)_0$}&
\colhead{$M_{\rm V}$}&
\colhead{$12+\rm{log(O/H)}$}&
\colhead{[Fe/H]}&
\colhead{$r_{\rm h}$}&
\colhead{References}\\
&
\colhead{(J2000)}&
\colhead{(J2000)}&
\colhead{(mag)}&
\colhead{(mag)}&
&
&
\colhead{(\arcmin)}&
}

\startdata
\multicolumn{9}{l}{-- Dwarf Spheroidals (dSph)\dotfill}\\

And\,XVIII& 00 02 14.5 & $+45$ 05 20 & $25.66\pm0.13$        &  $-9.7\pm0.1$ &        \nodata & $-1.80\pm0.10$ & $0.92\pm0.06$ & 1\\
   And\,XX& 00 07 30.7 & $+35$ 07 56 & $24.35^{+0.12}_{-0.15}$ &  $-6.3^{+1.1}_{-0.8}$ & \nodata & $-1.50\pm0.10$ & $0.53\pm0.14$ & 1,2\\
  And\,XIX& 00 19 32.1 & $+35$ 02 37 & $24.57^{+0.08}_{-0.36}$ &  $-9.2\pm0.6$ &        \nodata & $-1.90\pm0.10$ & $6.20\pm0.10$ & 1,2\\
     Cetus& 00 26 11.0 & $-11$ 02 40 & $24.39\pm0.07$        & $-11.2\pm0.2$ &        \nodata & $-1.90\pm0.10$ & $3.20\pm0.10$ & 1\\

  NGC\,147& 00 33 12.1 & $+48$ 30 32 & $24.15\pm0.09$        & $-14.6\pm0.1$ &        \nodata & $-1.10\pm0.10$ & $3.17$        & 1 \\
  And\,III& 00 35 33.8 & $+36$ 29 52 & $24.37\pm0.07$        & $-10.0\pm0.3$ &        \nodata & $-1.78\pm0.04$ & $2.20\pm0.20$ & 1\\
 And\,XVII& 00 37 07.0 & $+44$ 19 20 & $24.50\pm0.10$        &  $-8.7\pm0.4$ &        \nodata & $-1.90\pm0.20$ & $1.24\pm0.08$ & 1\\
  NGC\,185& 00 38 58.0 & $+48$ 20 15 & $23.95\pm0.09$        & $-14.8\pm0.1$ &  $8.20\pm0.20$ & $-1.30\pm0.10$ & $2.55$        & 1,3\\
    And\,I& 00 45 39.8 & $+38$ 02 28 & $24.36\pm0.07$        & $-11.7\pm0.1$ &        \nodata & $-1.45\pm0.04$ & $3.10\pm0.30$ & 1\\
   And\,XI& 00 46 20.0 & $+33$ 48 05 & $24.40^{+0.20}_{-0.50}$ &  $-6.9\pm1.3$ &        \nodata & $-2.00\pm0.20$ & $0.71\pm0.03$ & 1\\
  And\,XII& 00 47 27.0 & $+34$ 22 29 & $24.70\pm0.30$        &  $-6.4\pm1.2$ &        \nodata & $-2.10\pm0.20$ & $1.20\pm0.20$ & 1\\
  And\,XIV& 00 51 35.0 & $+29$ 41 49 & $24.33\pm0.33$        &  $-8.4\pm0.6$ &        \nodata & $-2.26\pm0.05$ & $1.70\pm0.80$ & 1\\
 And\,XIII& 00 51 51.0 & $+33$ 00 16 & $24.40^{+0.33}_{-0.40}$ &  $-6.7\pm1.3$ &        \nodata & $-1.90\pm0.20$ & $0.78\pm0.08$ & 1,2\\
   And\,IX& 00 52 53.0 & $+43$ 11 45 & $23.89^{+0.31}_{-0.08}$ &  $-8.1\pm1.1$ &        \nodata & $-2.20\pm0.20$ & $2.50\pm0.10$ & 1,2\\
  And\,XVI& 00 59 29.8 & $+32$ 22 36 & $23.60\pm0.20$        &  $-9.2\pm0.4$ &        \nodata & $-2.10\pm0.20$ & $0.89\pm0.05$ & 1\\
    And\,X& 01 06 33.7 & $+44$ 48 16 & $24.23\pm0.21$        &  $-7.6\pm1.0$ &        \nodata & $-1.93\pm0.11$ & $1.30\pm0.10$ & 1\\
    And\,V& 01 10 17.1 & $+47$ 37 41 & $24.44\pm0.08$        &  $-9.1\pm0.2$ &        \nodata & $-1.60\pm0.30$ & $1.40\pm0.20$ & 1\\
   And\,XV& 01 14 18.7 & $+38$ 07 03 & $24.00\pm0.20$        &  $-9.4\pm0.4$ &        \nodata & $-1.80\pm0.20$ & $1.21\pm0.05$ & 1\\
   And\,II& 01 16 29.8 & $+33$ 25 09 & $24.07\pm0.06$        & $-12.4\pm0.2$ &        \nodata & $-1.64\pm0.04$ & $6.20\pm0.20$ & 1\\
 And\,XXII& 01 27 40.0 & $+28$ 05 25 & $24.82^{+0.07}_{-0.31}$ & $-6.5\pm0.8$ &         \nodata & $-1.62\pm0.05$ & $0.94\pm0.10$ & 1,4,5\\
  Segue\,2& 02 19 16.0 & $+20$ 10 31 & $17.70\pm0.10$        &  $-2.5\pm0.3$ &        \nodata & $-2.00\pm0.25$ & $3.40\pm0.20$ & 1\\ 
   UMa\,II& 08 51 30.0 & $+63$ 07 48 & $17.50\pm0.30$        &  $-4.2\pm0.6$ &        \nodata & $-2.47\pm0.06$ & $16.0\pm1.0$  & 1\\
  Segue\,1& 10 07 04.0 & $+16$ 04 55 & $16.80\pm0.20$        &  $-1.5\pm0.8$ &        \nodata & $-2.72\pm0.40$ & $4.4^{+1.2}_{-0.6}$ & 1\\
Willman\,1& 10 49 21.0 & $+51$ 03 00 & $17.90\pm0.40$        &  $-2.7\pm0.8$ &        \nodata & $-2.10$        & $2.30\pm0.40$ & 1\\
    Leo\,V& 11 31 09.6 & $+02$ 13 12 & $21.25\pm0.12$        &  $-5.2\pm0.4$ &        \nodata & $-2.00\pm0.20$ & $2.60\pm0.60$ & 1\\
   Leo\,IV& 11 32 57.0 & $-00$ 32 00 & $20.94\pm0.09$        &  $-5.8\pm0.4$ &        \nodata & $-2.54\pm0.07$ & $4.60\pm0.80$ & 1\\
      Coma& 12 26 59.0 & $+23$ 54 15 & $18.20\pm0.20$        &  $-4.1\pm0.5$ &        \nodata & $-2.60\pm0.05$ & $6.00\pm0.60$ & 1\\
   CVn\,II& 12 57 10.0 & $+34$ 19 15 & $21.02\pm0.06$        &  $-4.9\pm0.5$ &        \nodata & $-2.20\pm0.05$ & $1.60\pm0.30$ & 1\\
Bootes\,II& 13 58 00.0 & $+12$ 51 00 & $18.10\pm0.06$        &  $-2.7\pm0.9$ &        \nodata & $-1.79\pm0.05$ & $4.20\pm1.40$ & 1\\
 Bootes\,I& 14 00 06.0 & $+14$ 30 00 & $19.11\pm0.08$        &  $-6.3\pm0.2$ &        \nodata & $-2.55\pm0.11$ & $12.6\pm1.0$  & 1\\
  Hercules& 16 31 02.0 & $+12$ 47 30 & $20.60\pm0.20$        &  $-6.6\pm0.4$ &        \nodata & $-2.41\pm0.04$ & $8.6^{+1.8}_{-1.1}$ & 1\\
  Segue\,3$\dagger$& 21 21 31.1 & $+19$ 07 03 & $16.1\pm0.1$ &  $-0.0\pm0.8$ &        \nodata & $-1.7^{+0.1}_{-0.3}$ & $0.47\pm0.13$ & 6\\
    Tucana& 22 41 49.6 & $-64$ 25 10 & $24.74\pm0.12$        &  $-9.5\pm0.2$ &        \nodata & $-1.95\pm0.15$ & $1.10\pm0.20$ & 1\\
Pisces\,II& 22 58 31.0 & $+05$ 57 09 & $21.31\pm0.18$        &  $-4.1\pm0.4$ &        \nodata & $-1.90$        & $1.10\pm0.10$ & 1,7 \\ 
  And\,VII& 23 26 31.7 & $+50$ 40 33 & $24.41\pm0.10$        & $-12.6\pm0.3$ &        \nodata & $-1.40\pm0.30$ & $3.50\pm0.10$ & 1\\
   And\,VI& 23 51 46.3 & $+24$ 34 57 & $24.47\pm0.07$        & $-11.3\pm0.2$ &        \nodata & $-1.30\pm0.14$ & $2.30\pm0.20$ & 1\\
  And\,XXI& 23 54 47.7 & $+42$ 28 15 & $24.67\pm0.13$        &  $-9.9\pm0.6$ &        \nodata & $-1.80\pm0.20$ & $3.50\pm0.30$ & 1\\

&&&&&&& \\

\multicolumn{9}{l}{-- Dwarf Irregulars (dIrr)\dotfill}\\

         WLM& 00 01 58.2& $-15$ 27 39 & $24.95\pm0.03$        & $-14.2\pm0.1$ &  $7.83\pm0.06$ & $-1.27\pm0.04$ & $7.78$        & 1,8,9\\
      IC\,10& 00 20 17.3& $+59$ 18 14 & $24.27\pm0.18$        & $-15.0\pm0.2$ &  $8.19\pm0.15$ & $-1.28$        & $2.65$        & 1,3,10\\
    IC\,1613& 01 04 47.8& $+02$ 07 04 & $24.39\pm0.12$        & $-15.2\pm0.2$ &  $7.62\pm0.05$ & $-1.60\pm0.20$ & $6.81$        & 1,8,11\\
      Leo\,A& 09 59 26.5& $+30$ 44 47 & $24.51\pm0.12$        & $-12.1\pm0.2$ &  $7.35\pm0.06$ & $-1.40\pm0.20$ & $2.15$        & 1,8,9,12\\
  Sextans\,B& 10 00 00.1& $+05$ 19 56 & $25.60\pm0.03$        & $-14.5\pm0.2$ &  $7.53\pm0.05$ & $-1.6$         & $1.06\pm0.10$ & 1,3,8,9\\
      Antlia& 10 04 04.1& $-27$ 19 52 & $25.65\pm0.10$        & $-10.4\pm0.2$ &        \nodata & $-1.60\pm0.10$ & $1.20\pm0.12$ & 1,13\\
  Sextans\,A& 10 11 00.8& $-04$ 41 34 & $25.60\pm0.03$        & $-14.3\pm0.1$ &  $7.54\pm0.06$ & $-1.85$        & $2.47$        & 1,8,9\\
    Sag\,DIG& 19 29 59.0& $-17$ 40 41 & $25.35\pm0.18$        & $-11.5\pm0.3$ &  $7.42\pm0.30$ & $-2.10\pm0.20$ & $0.91\pm0.05$ & 1,3,13\\
&&&&&&& \\                                                                                                                 

\multicolumn{9}{l}{-- Transition Dwarfs (dTrans or dIrr/dSph)\dotfill}\\

      LGS\,3& 01 03 55.0 & $+21$ 53 06 & $23.96^{+0.10}_{-0.07}$ & $-10.1\pm0.1$ &        \nodata & $-2.10\pm0.22$ & $2.10\pm0.20$ & 1,14\\
     Phoenix& 01 51 06.3 & $-44$ 26 41 & $23.09\pm0.10$        & $ -9.9\pm0.4$ &        \nodata & $-1.37\pm0.20$ & $3.76$        & 1,15\\
      Leo\,T& 09 34 53.4 & $+17$ 03 05 & $23.10\pm0.10$        &  $-8.0\pm0.5$ &        \nodata & $-1.99\pm0.05$ & $0.99\pm0.06$ & 1\\
    Aquarius& 20 46 51.8 & $-12$ 50 53 & $25.15\pm0.08$        & $-10.6\pm0.1$ &        \nodata & $-1.30\pm0.20$ & $1.47\pm0.04$ & 1,16\\
     Pegasus& 23 28 36.3 & $+14$ 44 35 & $24.82\pm0.07$        & $-12.2\pm0.2$ &  $7.93\pm0.13$ & $-1.40\pm0.20$ & $2.10$        & 1,8,9,16

\enddata 

\tablecomments{\ The half-light radius ($r_{\rm h}$) is the distance along the semimajor axis that contains half the light of the galaxy.}

\tablenotetext{$\dagger$}{\ Segue\,3 is likely a stellar cluster \citep[e.g.,][]{Belokurov+2010}.}

\tablerefs{\ Most values from (1) \citet{McConnachie+2012}, and
  references therein. Other references: (2) \citet{Watkins+2013}, (3)
  \citet{Mateo98}, (4) \citet{Martin+2009}, (5) \citet{Chapman+2013},
  (6) \citet{Fadely+2011}, (7) \citet{Sand+2012}, (8)
  \citet{HLee+2006}, (9) \citet{Tammann+2011}, (10) \citet{Kim+2009},
  (11) \citet{Bernard+2010}, (12) \citet{Bellazzini+2014}, (13)
  \citet{Pimbblet+2012}, (14) \citet{Miller+2001}, (15)
  \citet{Menzies+2008}, and (16) \citet{McConnachie+2005}.}

\end{deluxetable*}

\section{The Targets}
\label{sec:targets}

\subsection{Nearby Dwarf Galaxies}
We describe the DUSTiNGS targets and their properties in
Table~\ref{tab:targets} and Figure~\ref{fig:targs}. Dwarf galaxies are
the most prevalent morphological type of galaxy and may be the
building blocks of larger galactic systems
\citep{Tosi03}. Additionally, nearby dwarfs present a complete suite
of galactic environments \citep[e.g., metallicity and star formation
  history;][]{Mateo98,McConnachie+2012} that is perfect for studying
the connection between stellar populations and galaxy evolution.
DUSTiNGS includes all dwarf galaxies within 1.5~Mpc that were known at
the time of the observations and that lacked sufficient coverage with
{\it Spitzer} (see below). The next nearest galaxy ($d$=1.7~Mpc) is
beyond IRAC's ability to resolve stars. Following
\citet{McConnachie+2012}, we divide the nearby resolved dwarfs into
dwarf spheroidals (dSphs), dwarf irregulars (dIrrs), and transition
(dIrr/dSph, or dTrans) galaxies.

The dSphs typically have no detected neutral hydrogen and show no
evidence of recent star formation (within the last 200~Myr).  The dSph
galaxies are thought to have had their star formation terminated
either through an internal process such as a galactic wind
\citep[e.g.,][]{DekelSilk1986}, an external process such as an
interaction with a more massive host galaxy
\citep[e.g.,][]{Mayer+2001,Mayer+2006}, or heating by the ultraviolet
field associated with reionization
\citep[e.g.,][]{BabulRees1992,Efstathiou1992}.  

The dIrrs are gas rich and show evidence of \ion{H}{2} regions that
are sites of current massive star formation.  The dTrans galaxies
are typically gas rich, but show no evidence of current massive star
formation through the presence of \ion{H}{2} regions.  The nature of
transition galaxies is a matter of debate.  Many dTrans galaxies are
consistent with dIrr galaxies that are forming stars at such a low rate
that the absence of \ion{H}{2} regions is consistent with stochastic
variations. However, some show evidence for reduced gas mass fractions
and apparently lie between the dSphs and dIrrs in the
morphology-density relationship
\citep[e.g.,][]{Skillman+2003,Weisz+2011}.

Most of the DUSTiNGS galaxies are members of the Local Group
\citep{Mateo98,McConnachie+2012}. Based on their heliocentric radial
velocities, \citet{vandenBergh1999} argues that Sextans\,A, Sextans\,B,
and Antlia are not Local Group members, but instead belong to a
subgroup with NGC\,3109 that is expanding with the Hubble flow
\citep{vandenBergh1999}.

Of those known before our observations, we exclude fifteen galaxies
within 1.5 Mpc from DUSTiNGS because of existing {\it Spitzer}
observations. Nine of the most nearby dSph galaxies were observed in
cycle~5 using a similar observing strategy to the one employed
here (P.I.: P.\ Barmby, PID 50134: CVn\,I, Draco, Fornax, Leo\,I,
Leo\,II, Sculptor, Sextans, UMi, and UMa\,I).  Carina, NGC\,3109,
NGC\,6822, and NGC\,205 were also covered by several {\it
  Spitzer} programs (PIDs: 128, 159, 3126, 3400, 20469, 40204, 61001,
70062).  CMa and Sgr dSph are too large on the sky for efficient {\it
  Spitzer} imaging.  Because they are also nearby (7 and 26~kpc,
respectively), the Wide-Field Survey Explorer
\citep[WISE;][]{Wright+2010} all-sky IR survey is sufficiently
sensitive to detect a large fraction of the dust-producing stars.

\subsection{Expected Dusty Stellar Populations}
\label{sec:SFH}

Galaxies with different morphological types are expected to host
different sized AGB and massive star populations based on both the
typical mass scales and recent star formation histories.  For example,
dIrrs are typically more massive than the dSphs in the Local
Group, so they should have a larger population of dusty stars. However,
the level of recent star formation activity plays a significant role
in determining the number of these stars per unit stellar mass of a
galaxy. An intermediate-mass star enters the AGB stage of stellar
evolution between about 100 Myr and 3 Gyr after formation depending on
its initial mass \citep{Marigo+13}.  Thus, galaxies with higher rates
of star formation over these timescales will have larger populations
of AGB stars and galaxies with more recent star formation will have
massive stars. Because of the higher gas-rich content of dIrrs relative to
dSphs, the two factors of stellar mass and recent star formation
activity often compound one another.

\begin{figure}
  \includegraphics[width=\columnwidth]{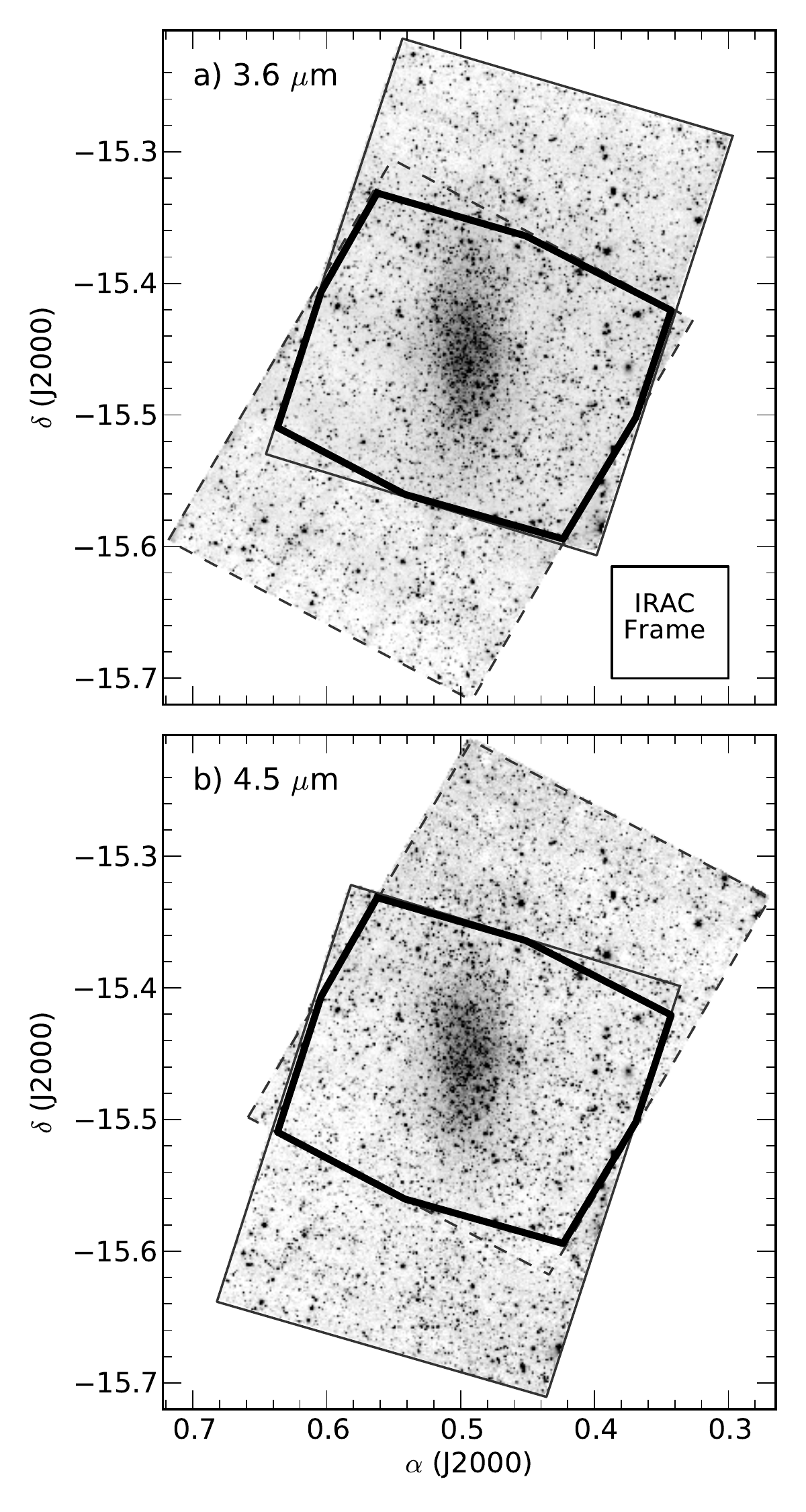}
\caption{DUSTiNGS mapping strategy. {\it a)} 3.6~\micron\ coadded
  mosaic for WLM. The solid thin line outlines the epoch~1 coverage
  and the dashed line outlines the epoch~2 coverage. The thick black
  line marks the coverage for all wavelengths and epochs, listed in 
  Table~\ref{tab:obs}. {\it b)} Same, for 4.5~\micron. A similar
  mapping scheme was implemented for every galaxy.  For WLM, the
  coverage is composed of a 3$\times$4 grid of IRAC frames
  (5\farcm2$\times$5\farcm2). Table~\ref{tab:obs} lists the grid size
  for each galaxy. The coverage listed in the last column of
  Table~\ref{tab:obs} includes only the area covered by all epochs and
  wavelengths. \label{fig:img}}
\end{figure}

However, differences do exist within each morphological type, with
dSphs showing the greatest divergence in recent star formation
activity \citep{Weisz+2014}, adding some uncertainty to expectations
on the AGB population from this morphological type. Detailed studies
of individual galaxies have shown that delayed onset of star formation
is also possible.  Both Leo~A \citep{Cole+2007} and Leo~T
\citep{Weisz+2012} are examples of gas-rich galaxies that have formed
the majority of their stars within the last 5-8 Gyr. Based on their
overall lower mass, the number of AGB stars in each of these
systems may be low even though a significant fraction of stellar mass in
each galaxy was formed over the timescale of interest.


\begin{deluxetable*}{lcccllllrrr}
\tablewidth{0pc}
\tabletypesize{\normalsize}
\tablecolumns{11}
\tablecaption{Data \& Observations\label{tab:obs}}

\tablehead{
\colhead{Galaxy}&
\colhead{$\langle$$t_{\rm exp}$$\rangle$\tablenotemark{a}}&
\colhead{5\,$\sigma$\tablenotemark{a}}&
\colhead{Map\tablenotemark{b}}&
\colhead{AOR Key}&
\colhead{Obs. Date}&
\colhead{AOR Key}&
\colhead{Obs. Date}&
\colhead{Separation}&
\colhead{$N_{\rm ptsrc}$\tablenotemark{c}}&
\colhead{Coverage\tablenotemark{d}}\\
&
\colhead{(s)}&
\colhead{($\mu$Jy)}&
\colhead{Size}&
&
\colhead{(UTC)}&
&
\colhead{(UTC)}&
\colhead{(days)}&
&
\colhead{(arcmin$^2$)}
}

\startdata
&&&&\multicolumn{2}{c}{--- Epoch 1 ---}&\multicolumn{2}{c}{--- Epoch 2 ---}&&\\
And\,I      &1080&1.6&2$\times$3&42307328& 2011 Sep 08 & 42307584& 2012 Mar 19 &193.1&  4640 &  85.6 \\
And\,II     &1080&1.6&2$\times$3&42307840& 2011 Sep 16 & 42308096& 2012 Mar 15 &181.3&  4309 &  85.4 \\
And\,III    &1080&1.6&2$\times$3&42308352& 2011 Sep 24 & 42308608& 2012 Mar 26 &184.7&  4043 &  85.9 \\
And\,V      &1080&1.6&2$\times$3&42309376& 2011 Sep 21 & 42309632& 2012 Mar 27 &188.0&  4877 &  85.8 \\
And\,VI     &1080&1.6&2$\times$3&42309888& 2011 Sep 24 & 42310144& 2012 Mar 09 &167.8&  4189 &  81.9 \\
And\,VII    &1080&1.6&2$\times$3&42310400& 2011 Aug 29 & 42310656& 2012 Mar 20 &203.4&  6951 &  79.9 \\
And\,IX     &1080&1.6&2$\times$3&42308864& 2011 Sep 23 & 42309120& 2012 Mar 27 &186.3&  4310 &  86.4 \\
And\,X      &1080&1.6&2$\times$3&42310912& 2011 Sep 19 & 42311168& 2012 Mar 17 &180.3&  4826 &  83.3 \\
And\,XI     &1080&1.6&2$\times$3&42311424& 2011 Sep 08 & 42311680& 2012 Mar 26 &200.4&  3200 &  83.9 \\
And\,XII    &1080&1.6&2$\times$3&42311936& 2011 Sep 24 & 42312192& 2012 Mar 27 &185.7&  3739 &  86.1 \\
And\,XIII   &1080&1.6&2$\times$3&42312448& 2011 Sep 21 & 42312704& 2012 Mar 27 &188.4&  3469 &  86.4 \\
And\,XIV    &1080&1.6&2$\times$3&42312960& 2011 Sep 19 & 42313216& 2012 Mar 21 &183.9&  3211 &  86.0 \\
And\,XV     &1080&1.6&2$\times$3&42313984& 2011 Sep 19 & 42314240& 2012 Mar 16 &178.9&  3794 &  84.5 \\
And\,XVI    &1080&1.6&2$\times$3&42314496& 2011 Sep 16 & 42314752& 2012 Mar 21 &187.3&  3164 &  86.5 \\
And\,XVII   &1080&1.6&2$\times$3&42315008& 2011 Sep 23 & 42315264& 2012 Mar 27 &186.1&  4736 &  86.2 \\
And\,XVIII  &1080&1.6&2$\times$3&42315520& 2011 Sep 06 & 42315776& 2012 Mar 17 &193.3&  4297 &  85.2 \\
And\,XIX    &1080&1.6&2$\times$3&42313472& 2011 Sep 24 & 42313728& 2012 Mar 17 &175.6&  3824 &  83.2 \\
And\,XX     &1080&1.6&2$\times$3&42316032& 2011 Aug 29 & 42316288& 2012 Mar 18 &201.7&  2992 &  82.8 \\
And\,XXI    &1080&1.6&2$\times$3&42329856& 2011 Sep 23 & 42330112& 2012 Mar 19 &178.2&  4505 &  82.8 \\
And\,XXII   &1080&1.6&2$\times$3&42330368& 2011 Sep 16 & 42330624& 2012 Mar 15 &181.2&  3121 &  85.6 \\
Antlia      &1080&1.6&2$\times$3&42316544& 2011 Jun 28 & 42316800& 2012 Feb 03 &219.6&  3666 &  76.6 \\
Aquarius    &1080&1.6&2$\times$3&42319616& 2011 Jun 22 & 42319872& 2012 Jan 06 &197.7&  3072 &  86.2 \\
Bootes\,I   &60  &9.1&4$\times$5&42317056& 2011 Sep 06 & 42317312& 2012 Mar 13 &189.2&  3249 & 354.9 \\
Bootes\,II  &60  &9.1&2$\times$3&42317568& 2011 Aug 28 & 42317824& 2012 Mar 13 &198.3&   850 &  79.5 \\
Cetus       &1080&1.6&2$\times$3&42318592& 2011 Sep 17 & 42318848& 2012 Feb 03 &139.7&  4041 &  79.8 \\
Coma        &60  &9.1&3$\times$4&42319104& 2011 Jul 18 & 42319360& 2012 Mar 13 &239.7&  1673 & 168.8 \\
CVn\,II     &150 &4.4&2$\times$3&42318080& 2011 Jul 26 & 42318336& 2012 Mar 13 &231.4&  2037 &  70.6 \\
Hercules\,Dw&150 &4.4&3$\times$4&42320640& 2011 Sep 20 & 42320896& 2012 Apr 24 &216.3&  5434 & 174.2 \\
IC\,10      &1080&1.6&3$\times$4&42321152& 2011 Sep 24 & 42321408& 2012 Apr 04 &193.1& 48057 & 195.9 \\
IC\,1613    &1080&1.6&4$\times$5&42321664& 2011 Sep 21 & 42321920& 2012 Feb 20 &153.2& 23538 & 356.3 \\
Leo\,A      &1080&1.6&2$\times$3&42322944& 2012 Jan 09 & 42322688& 2012 Jun 21 &164.0&  3680 &  83.1 \\
Leo\,IV     &150 &4.4&2$\times$3&42323200& 2011 Jul 18 & 42323456& 2012 Feb 15 &212.4&  1462 &  79.9 \\
Leo\,T      &1080&1.6&2$\times$3&42323968& 2012 Jan 08 & 42323712& 2012 Jun 21 &165.5&  3394 &  86.1 \\
Leo\,V      &150 &4.4&2$\times$3&42331392& 2011 Jul 17 & 42331648& 2012 Feb 15 &213.4&  1470 &  80.1 \\
LGS\,3      &1080&1.6&2$\times$3&42322176& 2011 Sep 21 & 42322432& 2012 Mar 19 &180.4&  2558 &  85.9 \\
NGC\,147    &1080&1.6&3$\times$4&42324224& 2011 Sep 23 & 42324480& 2012 Mar 30 &188.8& 33748 & 201.3 \\
NGC\,185    &1080&1.6&3$\times$4&42324736& 2011 Sep 19 & 42324992& 2012 Apr 04 &198.0& 32021 & 192.5 \\
Pegasus     &1080&1.6&3$\times$4&42320128& 2011 Sep 17 & 42320384& 2012 Jan 23 &127.5& 10688 & 179.8 \\
Phoenix     &1080&1.6&3$\times$4&42325248& 2011 Sep 09 & 42325504& 2012 Jan 19 &131.9&  9474 & 167.2 \\
Pisces\,II  &150 &4.4&2$\times$3&42331904& 2011 Aug 02 & 42332160& 2012 Jan 12 &163.6&  1205 &  77.9 \\
Sag\,DIG    &1080&1.6&2$\times$3&42326016& 2011 Nov 20 & 42325760& 2012 Jun 10 &202.3&  7102 &  85.8 \\
Segue\,1    &60  &9.1&2$\times$3&42326528& 2012 Feb 01 & 42326272& 2012 Jun 23 &142.7&   718 &  79.7 \\
Segue\,2    &60  &9.1&2$\times$3&42330880& 2011 Sep 23 & 42331136& 2012 Mar 15 &174.1&   598 &  79.8 \\
Segue\,3    &60  &9.1&2$\times$3&42332416& 2011 Jul 18 & 42332672& 2012 Jan 02 &167.7&  1048 &  75.1 \\
Sextans\,A  &1080&1.6&3$\times$4&42327040& 2012 Feb 01 & 42326784& 2012 Jul 19 &168.9&  8809 & 196.8 \\
Sextans\,B  &1080&1.6&3$\times$4&42327552& 2012 Feb 01 & 42327296& 2012 Jun 25 &145.4&  9631 & 195.7 \\
Tucana      &1080&1.6&2$\times$3&42327808& 2011 Jun 19 & 42328064& 2011 Nov 12 &146.1&  4374 &  72.5 \\
UMa\,II     &150 &4.4&3$\times$4&42328576& 2012 Jan 02 & 42328320& 2012 May 08 &126.8&  5056 & 163.1 \\
Willman\,1  &150 &4.4&2$\times$3&42329600& 2012 Jan 09 & 42329344& 2012 Jun 05 &148.4&  2321 &  70.6 \\
WLM         &1080&1.6&3$\times$4&42328832& 2011 Sep 10 & 42329088& 2012 Feb 01 &144.0& 12109 & 185.5

\enddata 

\tablenotetext{a}{\ The reported total exposure time per pixel and
  sensitivity are that of the combined epochs 1 and 2.}

\tablenotetext{b}{\ Map size is the number of frames on each axis.  A
  single IRAC frame is $5\farcm2 \times 5\farcm2$.}

\tablenotetext{c}{\ Total number of reliable point sources
  (Section~\ref{sec:phot}) within the spatial coverage listed in the
  last column.}

\tablenotetext{d}{\ Total coverage in arcmin$^2$ that is included at
  all epochs and all wavelengths (e.g., marked by the thick black line
  in Fig.~\ref{fig:img}).  This is smaller than the map size, which is
  the coverage at a single wavelength/epoch. This is the total
  coverage within which we can identify variable star candidates
  (Paper\,II). Galaxies with identical map sizes have slightly
  different total coverages owing to the rotation between the two
  epochs.}

\end{deluxetable*}


\section{Survey Design}
\label{sec:obs}

The DUSTiNGS survey includes uniform 3.6 and 4.5-\micron\ imaging of
50 nearby galaxies. These filters are particularly suited for
identifying sources with warm dust \citep[e.g., see the spectral
  energy distributions of dusty stars in Fig. 26
  from][]{Boyer+11}. The observations are summarized in
Table~\ref{tab:obs}. DUSTiNGS uses the InfraRed Array Camera
\citep[IRAC;][]{Fazio+04} onboard the {\it Spitzer Space Telescope}
\citep{Werner+04,Gehrz+2007} during the post-cryogen phase. The
spatial coverage extends to beyond the half-light radius ($r_{\rm h}$;
or the distance along the semimajor axis that contains half the
visible light of the galaxy) at each wavelength for determining the
level of foreground and background contaminating point sources. Each
galaxy was observed at two epochs approximately 6 months apart to
provide an additional diagnostic for identifying AGB stars, which are
variable at these wavelengths
\citep[e.g.,][]{LeBertre1992,LeBertre1993,McQuinn+2007,Vijh+09}. The
imaging footprint for WLM is shown in Figure~\ref{fig:img} as an
example of the DUSTiNGS mapping scheme.

Stellar evolution models \citep[e.g.,][]{Bressan+2012,Marigo+13} and
previous studies at these wavelengths
\citep[e.g.,][]{Jackson+07a,Jackson+07b,Boyer+09b} show that the tip
of the red giant branch (TRGB) is located at absolute
3.6-\micron\ magnitude $-6.6 \lesssim M_{3.6} \lesssim -6$~mag. Thus,
to ensure that the majority of thermally-pulsing (TP) AGB stars and
dust-producing massive stars would be detected, the exposure times
were chosen so that the 3-$\sigma$ detection limit is at least one
magnitude fainter than $M_{3.6}=-6$~mag. Together, the extended areal
coverage and sensitivity enable the detection of most of the
evolved stellar populations, thus significantly improving the
statistics on these short-lived evolutionary phases. In
particular, the DUSTiNGS sensitivity limit ensures the detection of
nearly all of the x-AGB stars; in the Magellanic Clouds, $>$96\% of
the x-AGB stars (Section~\ref{sec:xagb_class}) are brighter than
$M_{3.6}=-8$~mag.

For galaxies more distant than 400~kpc, we obtained 36 dithered frames
with 30~s exposures at each map position (deep observations,
$m_{3.6}^{5\sigma} \approx 20.5$~mag), one half of these frames were
obtained in each epoch.  Similarly, for galaxies with $130<d<400$~kpc,
we obtained 5 dithered frames with 30~s exposures (medium,
$m_{3.6}^{5\sigma} \approx 19.5$~mag) and for galaxies within 130~kpc,
we obtained 5 dithered frames with 12~s exposures (shallow,
$m_{3.6}^{5\sigma} \approx 18.5$~mag).  In each case, we used the
small cycling IRAC dither pattern with a median separation of
10.5~pixels to help eliminate imaging artifacts (the IRAC pixel size
is 1\farcs22). The map sizes and total exposure times ($t_{\rm exp}$)
are listed in Table~\ref{tab:obs}. The co-added, subsampled mosaics
are available for download at the Mikulski Archive for Space
Telescopes (MAST) and the InfraRed Science Archive (IRSA), and we show
examples in Figures~\ref{Afig:mos1} and \ref{Afig:mos2}.

\section{Point-source Photometry}
\label{sec:phot}

We describe below the photometry for the DUSTiNGS survey, including
the photometric corrections, saturation, completeness, and
crowding. The final photometric catalogs are available via MAST, IRSA,
and VizieR.

\subsection{PSF Photometry}
\label{sec:psf}

\begin{figure}
\vbox{
  \includegraphics[width=0.45\columnwidth]{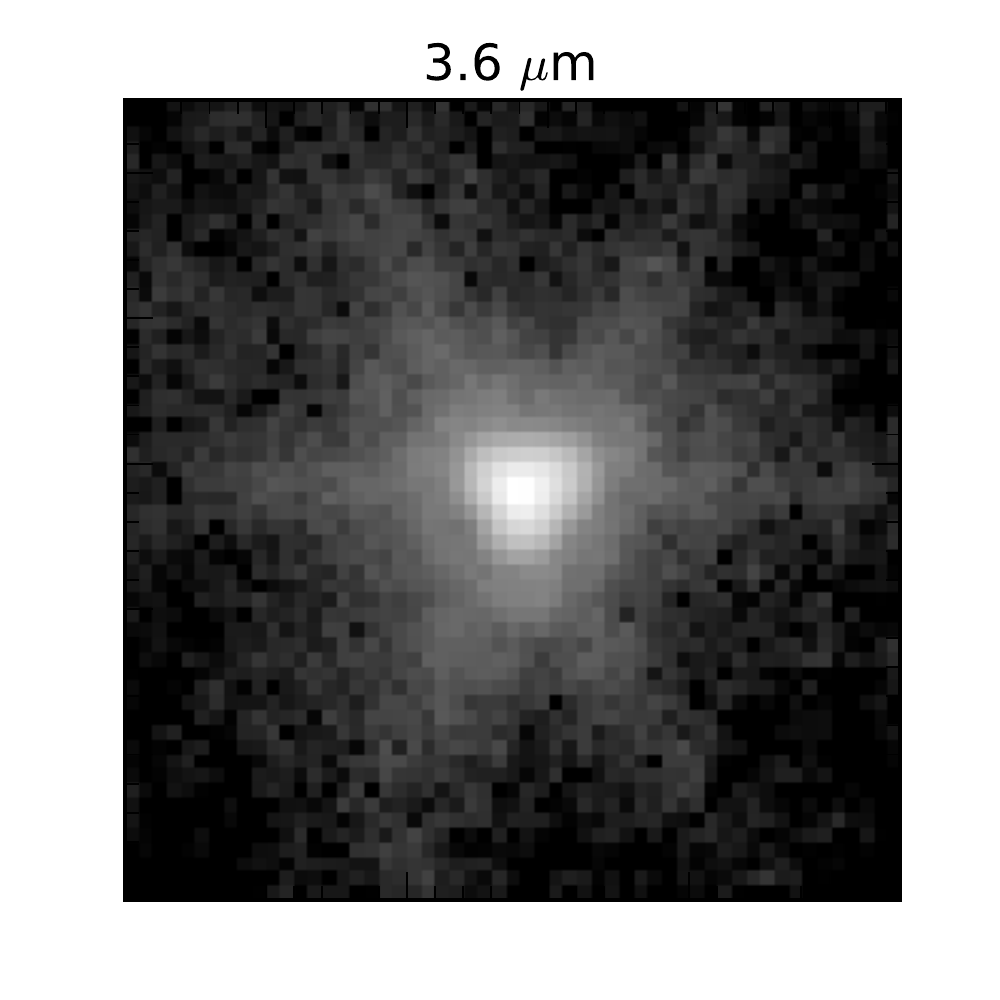}
  \includegraphics[width=0.45\columnwidth]{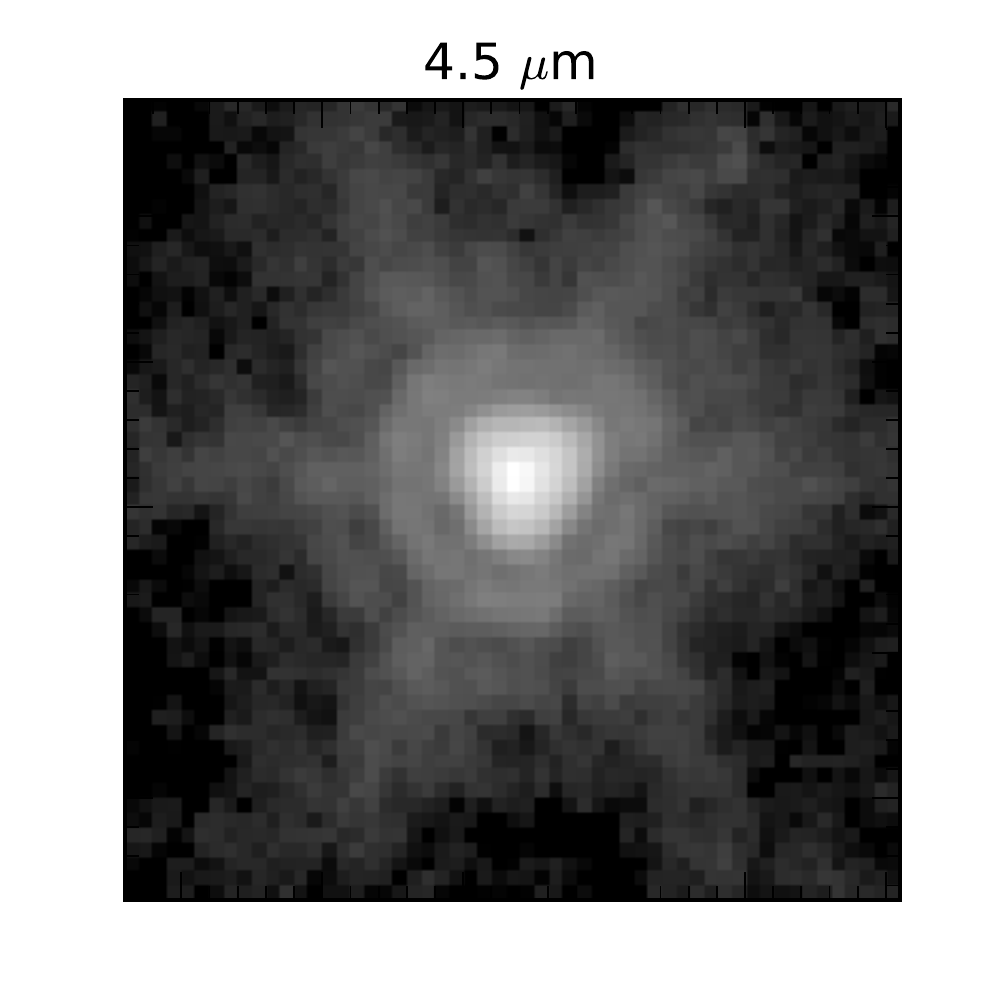}
}
\figcaption{The subsampled (0\farcs6 pixels) point-spread functions,
  constructed using the data in DAOphot. The images are scaled
  logarithmically to show the wing structure. \label{fig:psf}}
\end{figure}

Each galaxy in the DUSTiNGS survey was imaged over two epochs.  We
performed point-spread function (PSF) photometry separately for each
epoch to aid in identification of variable stars, and also for a
combined epoch to achieve the deepest photometry possible. Stars
brighter than $\approx$16~mag (see below) were measured on the
individual corrected Basic Calibrated Data (cBCD) frames from the {\it
  Spitzer} processing pipeline versions S18.18.0--S19.1.0 (depending
on the date of observations), using a weighted mean to combine the
measurements from each frame. The fainter magnitudes were recovered by
performing PSF photometry on the co-added cBCD frames, with sub-sampled
pixel sizes of 0\farcs6.  This two-step process is necessary to
achieve accurate photometry for both the faint and bright sources. The
photometry on the individual frames becomes unreliable at faint
magnitudes due to the Eddington bias \citep{Eddington1913}. This
effect causes stars to appear too bright when approaching the
detection limit because the source is more likely to be detected and
measured if random fluctuations on the detector make a source brighter
than its true flux.  On the other hand, bright sources are very
sensitive to the details of the PSF (Fig.~\ref{fig:psf}), so their
fluxes cannot be reliably measured on the mosaic where the PSF
features are smeared due to rotation between the frames. Fainter
sources are insensitive to these variations in the PSF and can thus be
accurately measured from the mosaic, allowing for the maximum
photometric depth.

All PSF photometry was carried out using DAOphot~II and {\sc ALLSTAR}
\citep{Stetson1987}, following a similar procedure to that used for
the Galactic Plane Survey Extraordinaire
\citep[GLIMPSE;][]{Benjamin+2003} and SAGE \citep{Meixner+06} programs
(B. Babler, private communication).  The PSF was constructed from the
data itself, using the Pegasus\,dIrr images to select $>$10 bright,
isolated stars with well-defined PSF wings (Fig.~\ref{fig:psf}). For
the photometry on the individual frames, we constructed the PSF using
a Moffat function \citep{Moffat1969} with $\beta=2.5$ for
3.6~\micron\ and $\beta=1.5$ for 4.5~\micron\, where a larger $\beta$
value approaches a Gaussian. The radius we used to fit the PSF to each
source was 1\farcs6--2\farcs0, or near the size of the FWHM. For the
mosaic photometry, the PSF is different because co-adding the images
smears the point sources. To achieve the best match between the cBCD
and mosaic photometry, we thus use a Moffat function ($\beta=1.5$) and
a Lorenz function for the 3.6 and 4.5~\micron\ mosaic PSFs,
respectively. The fitting radius was set to 3.2 pixels (1\farcs9).

In the final point-source catalog, the transition from cBCD to mosaic
photometry occurs at a magnitude where the photometry from both is
reliable and agrees to well within the photometric uncertainties. For
the medium and deep observations (150 s and 1080 s), this is at
16.5~mag and 15.7~mag for 3.6 and 4.5~\micron, respectively.  For the
shallow observations (60 s), the transition is at 15.5~mag and
15.0~mag, respectively. We note that there may be discontinuities in
the luminosity functions at the transition point.

\subsection{Photometric Corrections}
\label{sec:corr}

We applied several corrections to the photometry, as recommended by
the Spitzer Science Center (SSC). First, the cBCD images were
corrected for the pixel solid angle variation across the frame (at the
level of 1\%) and converted to data numbers for a robust measure of
the photometric uncertainties.

Second, sources were corrected for the variation in the point-source
flux across the array that is a result of the flat-fielding process
(the Array-Location-Dependent correction). This effect can be as high
as 10\%, depending on the location of the source within the
array. This correction is necessary for sources that are on the
Rayleigh-Jeans tail within the IRAC filters, which includes most of
the sources in our final catalogs.  Here, we do not apply the
Array-Location Dependent correction to point-sources that show a red
color ($[3.6]-[4.5] > 0$~mag) with a $>$3\,$\sigma$ significance.

Third, fluxes were adjusted by correcting for the location of the
center of the point-source within a pixel since the quantum efficiency
varies across each pixel (the Pixel Phase correction, up to
4\%). Fourth, we applied a color correction
for a 3000~K blackbody to the point-source fluxes, following the SSC's
recommendation.

\begin{figure}
\vbox{
  \includegraphics[width=\columnwidth]{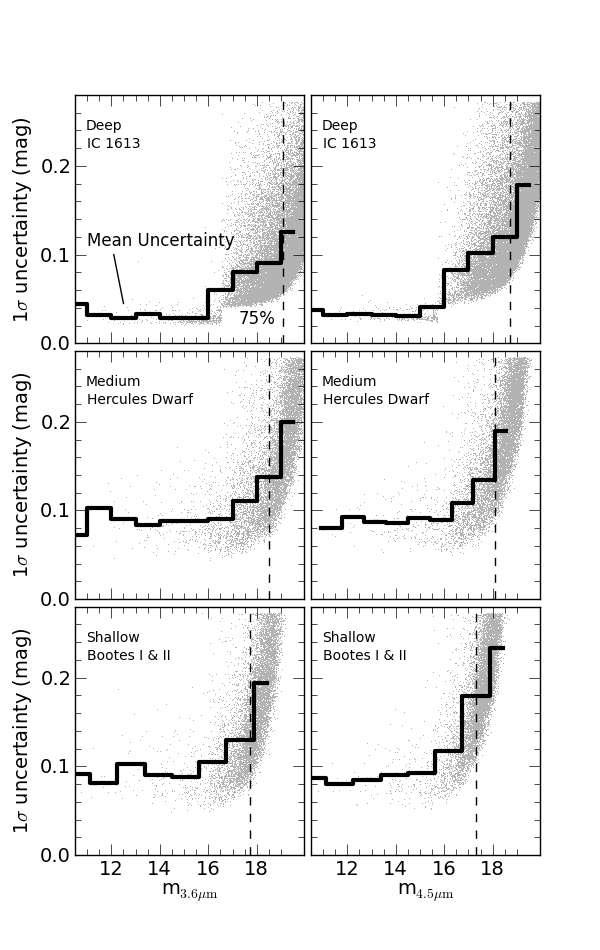}
} \figcaption{The photometric uncertainties from the Good-Source
  Catalog (GSC; Section~\ref{sec:cat}) for galaxies observed at all
  three total exposure times (Table~\ref{tab:obs}): IC\,1613 (top),
  Hercules Dwarf (middle), and both Bootes galaxies (bottom). The
  Bootes I and II galaxies are combined to illustrate the photometric
  uncertainties because they both have few point sources. The
  discontinuities near 16~mag are caused by the use of photometry on
  the cBCD frames for brighter sources and on the co-added frames for
  faint sources. The 75\% completeness level (Section~\ref{sec:comp})
  is shown as a dashed line in each panel. The histogram indicates the
  mean uncertainty at a given magnitude. \label{fig:errors}}
\end{figure}

Following \citet{FruchterHook2002}, we increased the measured
flux uncertainties by a factor of 2 to account for correlated
uncertainties between the pixels that arise from subsampling the
mosaic. This correction was only applied to sources measured from the
mosaics. Along with this uncertainty, the final photometric
uncertainties include those reported by DAOphot and the calibration
uncertainties listed by \citet{Reach+05} (Fig.~\ref{fig:errors}).

The DUSTiNGS catalog includes magnitudes using the Vega-based zero
points of $280.9\pm4.1$~Jy for 3.6~\micron\ and $179.7\pm2.6$~Jy for
4.5~\micron. The final photometry is well
matched to that from WISE, which has filters similar to IRAC (3.4 and
4.6~\micron, or W1 and W2).  Agreement is within 0.02~mag down to the
repeatability limit of the WISE photometry
($\approx$14~mag). IRAC
point-source positions are accurate to
$\approx$0\farcs5.

\subsection{Saturation}
\label{sec:sat}

\begin{figure}
  \includegraphics[width=\columnwidth]{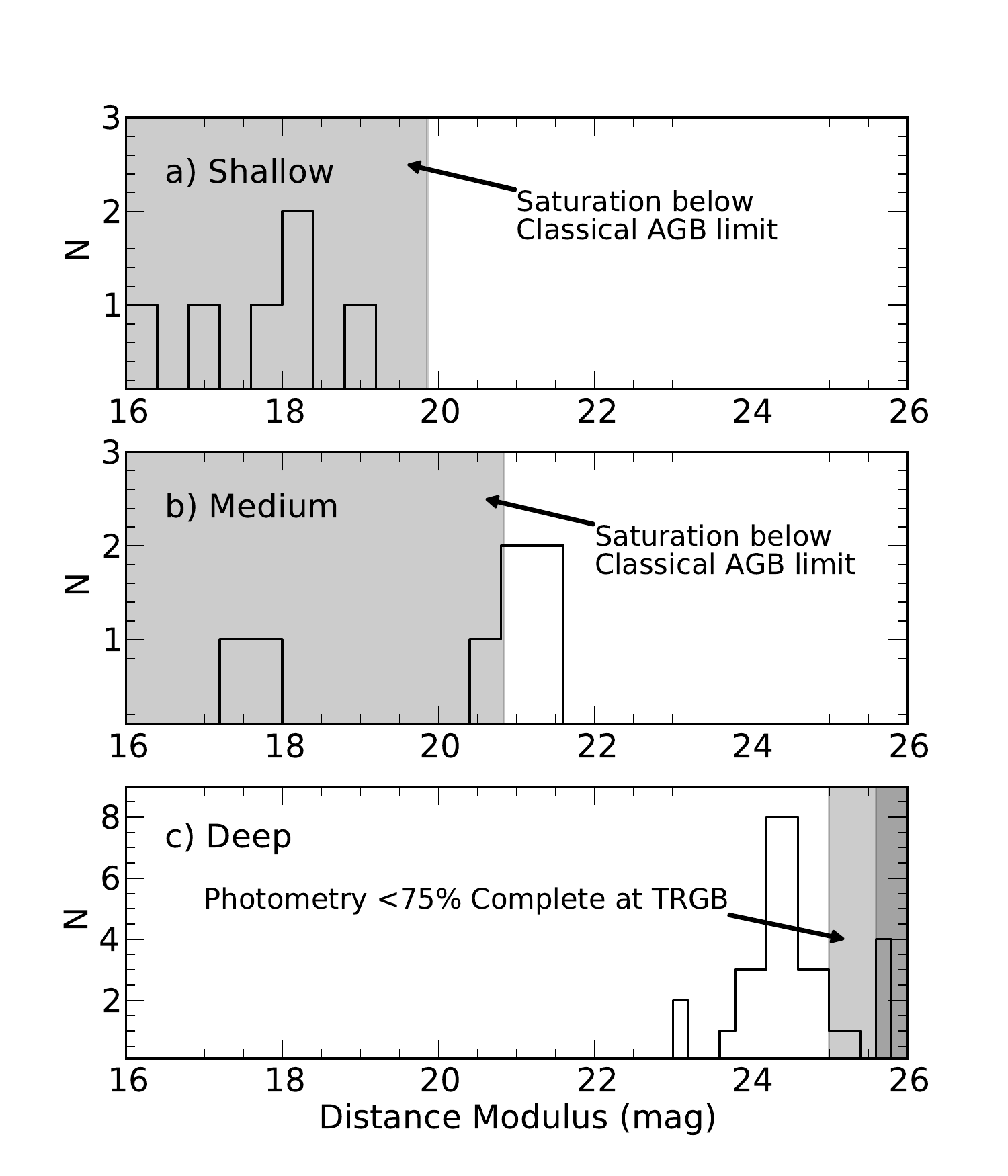}
\caption{Number distribution of galaxies with saturation and
  photometric incompleteness for 3.6~\micron. {\it a)} Galaxies with
  60~s total integrations (12~s per frame). The shaded region marks
  the distance moduli where some stars fainter than the classical AGB
  limit ($M_{3.6} \gtrsim -10$~mag) will saturate. Bright (massive)
  stars in these galaxies are at risk of saturating. {\it b)}
  Galaxies with 150~s total integrations (30~s per frame). Those
  within the shaded region are at risk of saturating the brightest AGB
  stars. {\it c)} Galaxies with the deepest (1080~s) integrations
  (30~s per frame). AGB stars within these galaxies are not at risk of
  saturation. However, those with distance moduli within the light and
  dark shaded regions have $<$75\% photometric completeness at
  $M_{3.6} = -6$~mag and $-6.6$~mag, respectively, which is the
  assumed range of the TRGB.\label{fig:sat}}
\end{figure}

The saturation limits for 30~s frames are 10.84~mag and 10.35~mag for
3.6 and 4.5~\micron, respectively. For 12~s frames, the saturation
limits are 9.86~mag and 9.34~mag. For the galaxies with $(m-M)_0 >
23$~mag, saturation only occurs for stars $\gtrsim$2~mag brighter than
the classical AGB limit, which lies near $-10 > M_{3.6} > -11$~mag
\citep[$M_{\rm bol} = -7.1$~mag, derived for 3.6~\micron\ using the
  models from][]{Groenewegen06}. This includes all of the dIrr and
dTrans galaxies.  Because these galaxies are more likely to show
evidence of recent/ongoing star formation, they are more likely to
include massive AGB stars which approach (and sometimes slightly
exceed) the classical AGB limit.

The nearest galaxies with $(m-M)_0 < 22$~mag are all dSph galaxies
with little to no ongoing star formation.  Any dust-producing stars in
these galaxies are thus more likely to have low initial masses and
luminosities near the TRGB.  Nevertheless, saturation
does occur at magnitudes fainter than the classical AGB limit for 9
galaxies (those within the shaded regions in Fig.~\ref{fig:sat}a,b):
Bootes\,I, Bootes\,II, Coma, Hercules, Segue\,1, Segue\,2, Segue\,3,
UMa\,II, and Willman\,1.

\subsection{Photometric Completeness}
\label{sec:comp}

\begin{figure}
\vbox{
  \includegraphics[width=\columnwidth]{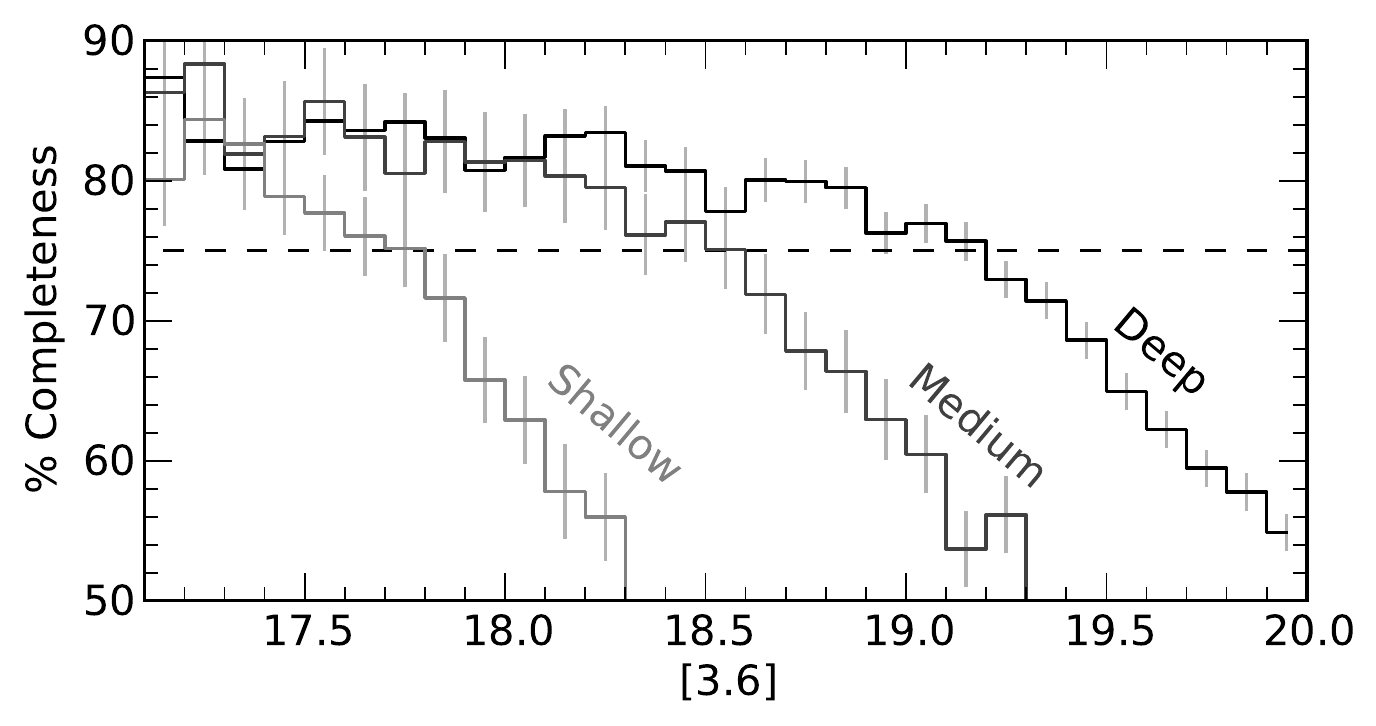}
\includegraphics[width=\columnwidth]{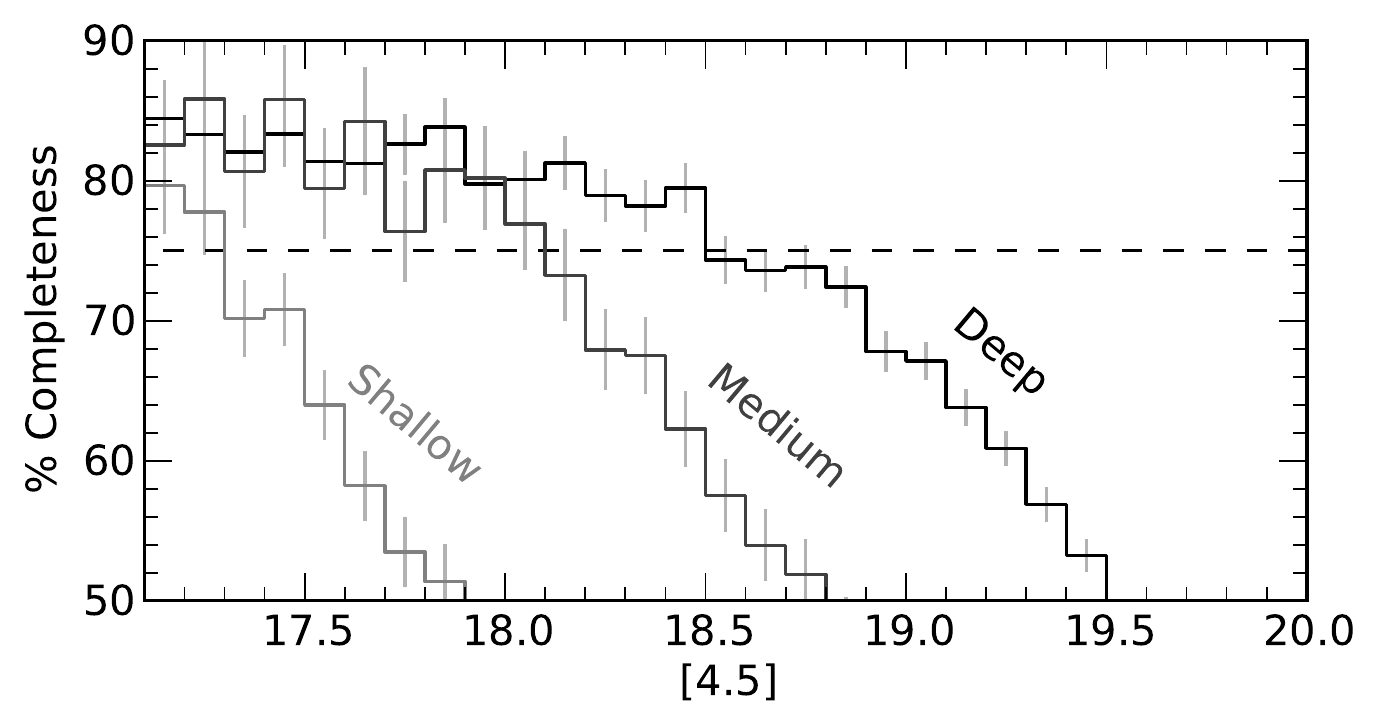}
}
\caption{Average completeness curves for the shallow, medium, and deep
  epoch\,1 data. For the deeper, combined-epoch photometry, the
  completeness limits are approximately 0.5~mag fainter. The dashed
  line marks 75\% completeness. These curves were derived for off
  regions (Section~\ref{sec:comp}) and reflect only completeness due
  to sensitivity. These curves exclude galaxies that suffer from
  additional crowding; for those galaxies, the curves have similar
  shapes, shifted towards the brighter magnitudes listed in
  Table~\ref{tab:comp}. \label{fig:complim}}
\end{figure}

\begin{figure}
  \includegraphics[width=\columnwidth]{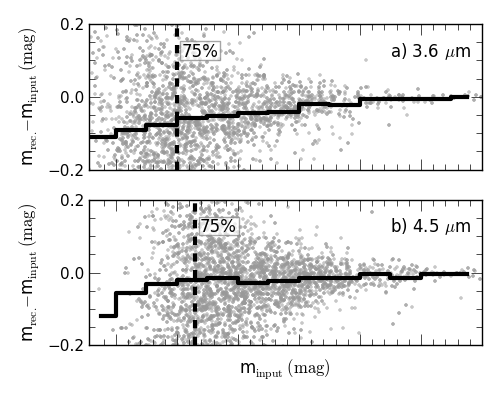}
\caption{The difference between the input stellar magnitudes and the
  recovered stellar magnitudes from the artificial star tests.  Three
  galaxies with $t_{\rm exp} = 1080$~s are shown here. The 75\%
  completeness limit is shown as a dashed line, and the solid black
  line shows the mean magnitude difference within 0.5~mag bins,
  excluding sources outside 3\,$\sigma$. The magnitudes in the
  published catalogs are corrected for the bias shown
  here. \label{fig:false}}
\end{figure}

To assess the repeatability of the photometry, we performed artificial
star tests. For each galaxy and wavelength, we added 20 artificial
stars of varying magnitudes to a 25 square arcmin region that excludes
the galaxy center (crowding in the galaxy centers is discussed in
Section~\ref{sec:crowd}).  This was repeated 100 times, for a total of
2000 artificial stars. The magnitude distribution of the fake stars
mimicked the real magnitude distribution (see Fig.~\ref{fig:false}).
Table~\ref{tab:comp} lists the mean and the standard deviation in the
resulting photometric completeness limits for galaxies unaffected by
crowding (Section~\ref{sec:crowd}) and the completeness curves
  are shown in Figure~\ref{fig:complim}.

The mean difference between the magnitudes of the added and recovered
stars shows a small bias that increases with magnitude, but is
$\lesssim$0.06~mag and 0.02~mag for 3.6 and 4.5~\micron, respectively,
for stars brighter than the 75\% completeness limit
(Fig.~\ref{fig:false}). For stars near 20$^{\rm th}$ magnitude, the
mean difference is $\lesssim$0.1~mag. This bias is consistent
  with the effects of point-source crowding, which biases measurements
  towards brighter magnitudes and increases for faint sources. While
  only a few galaxies are affected by crowding above the TRGB
  (Section~\ref{sec:crowd}), all DUSTiNGS galaxies are affected by
  crowding at faint magnitudes. The final magnitudes are corrected
for this bias.

For most galaxies, the photometry is better than 75\% complete at
$M_{3.6}=-6$~mag in each epoch, which is the approximate faint limit
for the TRGB \citep{Jackson+07a,Boyer+09b,Boyer+11}. At brighter
magnitudes, the completeness rapidly increases
(Fig.~\ref{fig:complim}); we report the 75\% limit throughout this
work because it is representative of the completeness level near the
TRGB for the most distant target galaxies. Six DUSTiNGS galaxies have
$<$75\% complete photometry at $-6$~mag (light shaded region of
Fig.~\ref{fig:sat}c) in a single epoch, though all six reach 75\%
completeness by $-6.7$~mag. In the photometry from the combined
epochs, the completeness limit is approximately 0.5~mag fainter,
resulting in near-complete photometry to the TRGB in all 50 galaxies.


\begin{deluxetable}{rcc}
\tablewidth{\columnwidth}
\tabletypesize{\normalsize}
\tablecolumns{3}
\tablecaption{75\% Photometric Completeness Limits\label{tab:comp}}

\tablehead{
\colhead{$\langle t_{\rm exp} \rangle$}&
\colhead{3.6\,\micron}&
\colhead{4.5\,\micron}\\
\colhead{(s)}&
\colhead{(mag)}&
\colhead{(mag)}
}

\startdata
60   & $17.7 \pm 0.2$ & $17.3 \pm 0.1$\\
150  & $18.5 \pm 0.2$ & $18.1 \pm 0.1$\\
1080 & $19.1 \pm 0.1$ & $18.7 \pm 0.2$\\
\hline
\noalign{\vskip 1mm}
\multicolumn{3}{c}{-- Galaxies affected by extrinsic crowding\tablenotemark{a} --} \\
And\,VII  & $18.5 \pm 0.2$ & $18.5 \pm 0.2$ \\
IC\,10    & $17.7 \pm 0.2$ & $17.7 \pm 0.2$ \\
NGC\,147  & $18.2 \pm 0.2$ & $18.3 \pm 0.2$ \\
NGC\,185  & $18.6 \pm 0.2$ & $18.4 \pm 0.2$ \\
Sag\,DIG  & $18.2 \pm 0.2$ & $17.9 \pm 0.2$ 
\enddata

\tablenotetext{a}{\ All galaxies affected by extrinsic crowding have
  $\langle t_{\rm exp} \rangle = 1080$~s.}

\tablecomments{\ Completeness limits ($m_{75\%}$) were computed for
  the epoch 1 data. For the deeper, combined-epoch photometry, the
  completeness limit is approximately 0.5~mag fainter. The first three
  rows list the mean and standard deviation of the completeness limit
  for galaxies unaffected by crowding. All limits in this table were
  derived from a 25~arcmin$^2$ region away from the galaxy's center.}

\end{deluxetable}


\subsection{Crowding}
\label{sec:crowd}

Stellar crowding affects the photometric completeness both in the
centers of dense galaxies (intrinsic) and for galaxies near the
Galactic Plane, where foreground stars from the Milky Way increase the
stellar density (extrinsic; Fig.~\ref{fig:targs}).

We compute the photometric completeness as a function of radius to
measure crowding from stars within the galaxies themselves.  For most
DUSTiNGS galaxies, internal crowding does not significantly affect the
photometry. WLM and Sextans\,A show only slight crowding within
1\arcmin\ of the their centers, affecting the photometric completeness
by $\lesssim$0.2~mag at 3.6~\micron.  Severe crowding is evident for
IC\,10, NGC\,147, and NGC\,185.  Table~\ref{tab:crowd} lists the
radius where the photometry becomes 75\% complete at absolute
magnitudes of $M_{3.6}=-6$ and $-8$~mag, which are the limits used to
identify AGB candidates in Section~\ref{sec:pops}. IC\,10 is the only
galaxy for which the number of x-AGB (Section~\ref{sec:agb}) candidates
should be considered a lower limit.

All galaxies residing well above or below the Galactic Plane show
similar completeness limits, but the 75\% completeness limit rapidly
increases in brightness as the distance from the Galactic Plane
decreases. IC\,10 and And\,VII have the smallest Galactic latitudes
and are the most affected by foreground stars (Fig.~\ref{fig:targs}).
Sag\,DIG has a higher Galactic latitude but its longitude places it
near the Galactic Bulge. The completeness limits for galaxies affected
by extrinsic crowding are listed in Table~\ref{tab:comp}.

\begin{deluxetable}{cccc}
\tablewidth{\columnwidth}
\tabletypesize{\normalsize}
\tablecolumns{4}
\tablecaption{Intrinsic Crowding Limits\label{tab:crowd}}

\tablehead{
\colhead{Galaxy}&
\colhead{$m_{3.6}$}&
\colhead{$M_{3.6}$}&
\colhead{$M_{3.6}$}\\
\colhead{}&
\colhead{$=m_{75\%}$}&
\colhead{$=-6$\,mag}&
\colhead{$=-8$\,mag}
}

\startdata
IC\,10   & $R\approx5\arcmin$ & \nodata  & $R\approx1\arcmin$ \\
NGC\,147 & $R\approx4\arcmin$ & $R\approx4\arcmin$ & All $R$ \\
NGC\,185 & $R\approx4\arcmin$ & $R\approx3\arcmin$ & All $R$ 
\enddata

\tablecomments{\ The radii ($R$) where 75\% completeness is reached
  for the given absolute magnitudes in galaxies that suffer from
  intrinsic crowding in their centers. Crowding was measured at radius intervals
  of 1\arcmin. Note that NGC\,147 is elongated \citep[ellipticity $
    \epsilon = 0.41\pm0.02$;][]{McConnachie+2012}, so photometry is
  complete at smaller radii along the minor axis.}

\end{deluxetable}

\section{Description of the Catalog}
\label{sec:cat}

\begin{figure}
  \includegraphics[width=\columnwidth]{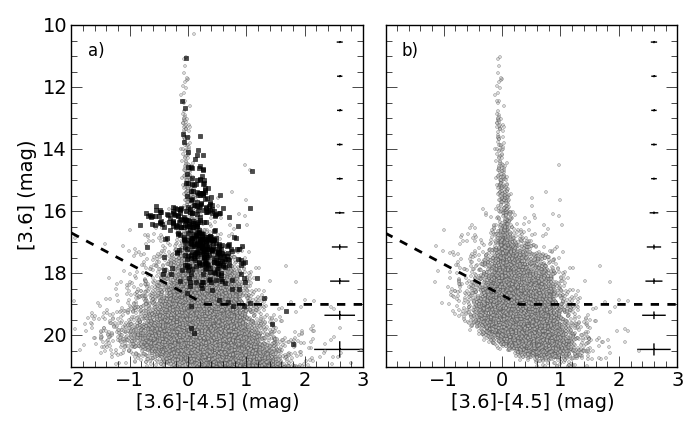}
\caption{CMDs for
  And\,I showing {\it (a)} the full catalog and {\it (b):} the
  good-source catalog. In the full catalog, marginally resolved
  sources are marked by darker points (Section~\ref{sec:gal}). The
  $[3.6]-[4.5]$ colors of extended sources can be
  artificially red or blue because fluxes may be extracted from the
  individual frames for one wavelength and from the mosaics for the
  other wavelength.  The dashed line marks the 75\% completeness
  limit. Mean photometric uncertainties are shown on the right of each
  panel. \label{fig:gsc}}
\end{figure}

The final Vega magnitudes of the high-quality point sources are
reported in the DUSTiNGS ``Good''-Source Catalog (GSC), which is
described in Table~\ref{tab:cat} and is available to download from
MAST, VizieR, and
IRSA. To construct the GSC, we culled the
full photometric catalogs using the sharpness ($S$) and chi ($\chi$)
parameters returned by DAOphot. To eliminate artifacts and extended
objects, the sharpness value is restricted to $-0.3 < \langle
S_\lambda \rangle < 0.3$. The $\chi$ parameter is a measure of the
root-mean-square of the residuals and is restricted to $\langle
\chi_\lambda \rangle < 5$ for sources measured from the cBCD frames
and to $\langle \chi_\lambda \rangle < 2$ for those measured from the
mosaics. In addition, the GSC includes only sources detected above the
4\,$\sigma$ level and below the saturation limit and is restricted to
sources that meet these criteria at both 3.6 and 4.5~\micron.
Figure~\ref{fig:gsc} shows an example color-magnitude diagram (CMD) of
the full catalog compared with the GSC. The CMDs using the GSC for all
targeted galaxies are presented in Figures~\ref{Afig:cmds1} and
\ref{Afig:cmds2}.

\begin{deluxetable}{ll}
\tablewidth{\columnwidth}
\tabletypesize{\footnotesize}
\tablecolumns{2}
\tablecaption{GSC Catalog Description\label{tab:cat}}

\tablehead{
\colhead{Column}&
\colhead{Description}
}

\startdata
1      & Galaxy Name \\
2      & Point-source name; IAU convention\\
3--4   & RA (h:m:s), Dec ($\degr:\arcmin:\arcsec$); J(2000)\\
5--8   & 3.6~\micron\ mag and uncertainty for Epochs 1 \& 2 \\
9--12  & 4.5~\micron\ mag and uncertainty for Epochs 1 \& 2 \\
13--14 & DAOphot $S_{3.6}$ values for Epochs 1 \& 2 \\
15--16 & DAOphot $S_{4.5}$ values for Epochs 1 \& 2 \\
17--18 & DAOphot $\chi_{3.6}$ values for Epochs 1 \& 2 \\
18--19 & DAOphot $\chi_{4.5}$ values for Epochs 1 \& 2 

\enddata 

\tablecomments{\ The catalog is available for download via MAST, IRSA, and VizieR.}
\end{deluxetable}


\subsection{Marginally Resolved Extended Sources}
\label{sec:gal}

Extended sources that are unresolved or marginally
resolved in the individual cBCD frames are more strongly resolved in
the subsampled mosaic. There are several sources measured from the
cBCD frames that therefore meet the sharpness criteria for the GSC,
but would fail the same criteria if measured on the
mosaic. Because these sources are extended, PSF photometry
  is inappropriate and can result in large uncertainties; the
  PSF-derived magnitude measured on the cBCD frames can differ from
  that measured from on mosaic by 0.2 to 1~mag.  For sources near
the transition magnitude where the individual-frame photometry and the
mosaic photometry were combined (Section~\ref{sec:psf}), this results
in artificially blue or red colors (dark points in Fig.~\ref{fig:gsc})
if stars were measured on the cBCD frames for one wavelength and on
the mosaics for the other. These sources are easily identified via a
mean sharpness value created by combining $S_\lambda$ measured by
DAOphot for all measured channels and epochs from both the cBCD frames
and mosaics. This combined sharpness parameter is larger for
marginally resolved sources than for the true point-sources at a given
magnitude.

Removing these sources from the GSC significantly decreases the
contamination from background sources brighter than $\approx$17~mag
and allows for a more accurate selection of stars belonging to the
target galaxies. We do not remove these sources from the full catalog
because we cannot rule out the possibility that they are indeed galaxy
members (e.g., star clusters). However, the PSF-derived
  magnitudes for these sources are unreliable, so we include only
  their positions in the full catalog and recommend aperture
  photometry for accurate fluxes.

\subsection{Extinction}
\label{sec:ext}

We have not corrected for extinction in the photometric catalogs. With
the exception of IC\,10, all DUSTiNGS galaxies show E$(B-V) < 0.2$~mag
\citep{McConnachie+2012}. At 3.6 and 4.5~\micron, this level of
extinction results in a change in magnitude that is less than the
photometric uncertainties ($A_{3.6}<0.03$~mag and $A_{4.5}<0.02$~mag).

IC\,10 has the smallest Galactic latitude, and thus the highest level
of extinction at E$(B-V) = 1.6$~mag. Correcting the IRAC magnitudes
for extinction would result in a magnitude decrease of $\sim$0.2~mag. However,
the change in color due to extinction is still well below the photometric
uncertainties with $\Delta$$(m_{3.6}-m_{4.5}) < 0.04$~mag.

\begin{figure}
  \includegraphics[width=\columnwidth]{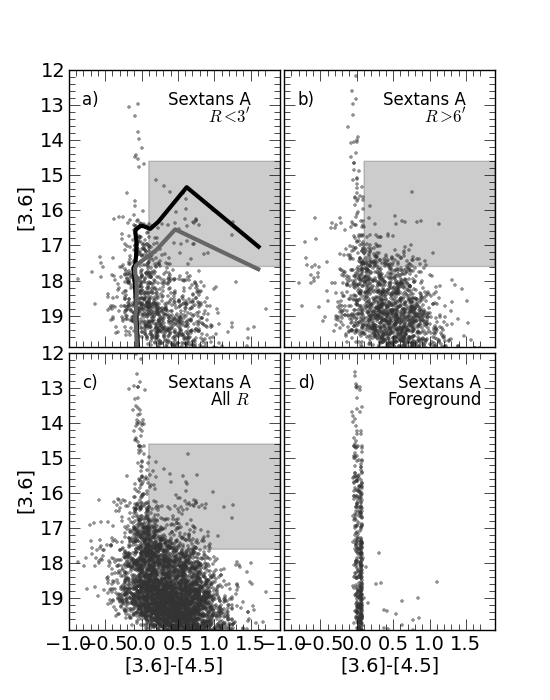}
\caption{Epoch 1 CMD for Sextans\,A, showing {\it a)} sources within
  3\arcmin, {\it b)} sources beyond 6\arcmin, {\it c)} the entire
  coverage, and {\it d)} foreground simulation for the full spatial
  coverage (Table~\ref{tab:obs}) from TRILEGAL. The shaded region
  shows the approximate location of x-AGB stars, based on their
  position on the same CMD in the Magellanic Clouds
  \citep[Section~\ref{sec:xagb_class};][]{Blum+06,Bolatto+07,Boyer+11}.  The
  half-light radius for Sextans\,A is 2\farcm47, with an ellipticity
  of only 0.17 \citep[Fig.~\ref{Afig:mos2};][]{McConnachie+2012}. In
  panel {\it a)}, the dark and light solid lines are 400~Myr and 1~Gyr
  isochrones, respectively, from \citet{Marigo+2008}. \label{fig:fg}}
\end{figure}

\subsection{Background and Foreground Contamination}
\label{sec:cmd}

The DUSTiNGS field-of-view is large enough to provide a robust
estimate of the foreground and background
sources. Figure~\ref{fig:fg}c shows the epoch\,1 CMD for Sextans\,A,
one of the more distant DUSTiNGS galaxies ($r_{\rm h} = 2\farcm47$;
also see Fig.~\ref{Afig:mos2}). To demonstrate a CMD
  with minimal contamination from nonmembers and a CMD that is
  dominated by background and foreground, we also show the CMDs of
  inner and outer regions of the Sextans\,A coverage in
  Figure~\ref{fig:fg}a,b. We show an estimate of the foreground in
  panel {\it d}, simulated with the TRILEGAL stellar population
  synthesis code \citep{Girardi+05}.  The difficulty in
distinguishing between dusty stars with $[3.6]-[4.5]>0.1$~mag and
$M_{3.6} < -8$~mag and unresolved background sources in the same
color-magnitude space (shaded region of Fig.~\ref{fig:fg}) is clear
when comparing panels {\it a)} and {\it b)}. Less dusty member stars
(with $[3.6]-[4.5] \approx 0$~mag) are also difficult to identify due
to confusion with both background and foreground sources. Because AGB
stars and some massive stars are variable, the dual-epoch DUSTiNGS
observations are crucial for identifying individual member stars
(Paper\,II).

In galaxies with a large intermediate-aged stellar population, a
branch of x-AGB stars (Section~\ref{sec:xagb_class}) that follows the
isochrones shown in Figure~\ref{fig:fg}a is easily identifiable in the
CMD (Figs.~\ref{Afig:cmds1} and \ref{Afig:cmds2}). This feature is
clearly visible in only a handful of the DUSTiNGS galaxies: IC\,10,
IC\,1613, NGC\,147, and NGC\,185. Even in other star-forming DUSTiNGS
galaxies (e.g., WLM, Sag\,DIG, Sextans\,A, Sextans\,B, and
Pegasus\,dIrr), this branch is not easily distinguished from
background sources.

\subsection{Luminosity Functions}

\begin{figure}
  \includegraphics[width=\columnwidth]{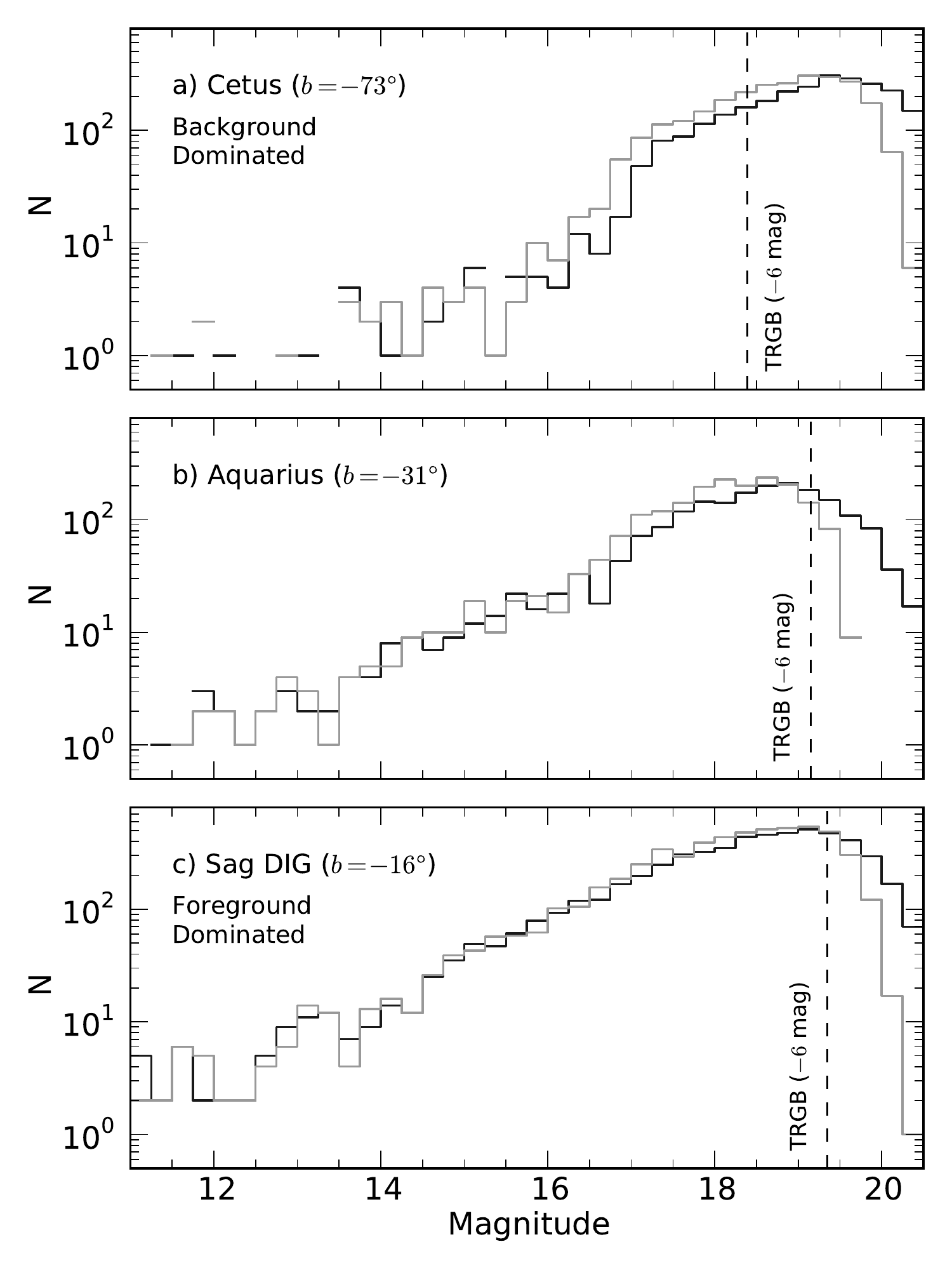}
\caption{Luminosity functions for {\it a)} Cetus, {\it b)} Aquarius,
  and {\it c)} Sag\,DIG.  The black and gray lines are the
  3.6~\micron\ and 4.5~\micron\ luminosity functions,
  respectively. Because Cetus lies far from the Galactic Plane, its
  luminosity function is dominated by red background sources, mostly
  fainter than 17~mag. Sag\,DIG ($b=-16\degr$) is dominated by
  foreground from the Galactic Bulge. In all panels, the expected
  TRGB ($M_{\rm [3.6]} \approx -6$~mag) is marked with a dashed
  line. \label{fig:lfunc}}
\end{figure}

\begin{figure}
  \includegraphics[width=\columnwidth]{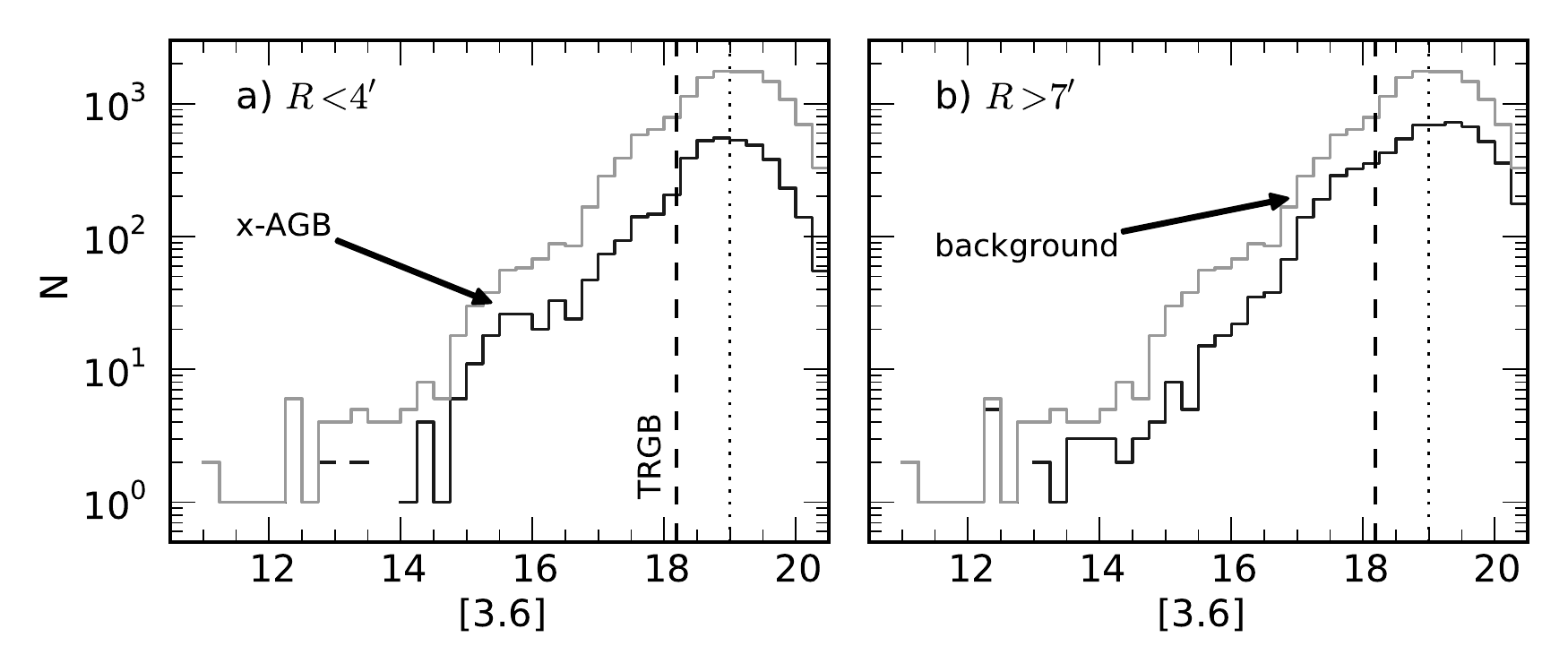}
\caption{3.6~\micron\ luminosity function for IC\,1613 for {\it a)} an
  on region ($r_{\rm h} = 6\farcm8$), and {\it b)} an off region. The
  dusty AGB stars (x-AGB; Section~\ref{sec:xagb_class}) are clearly visible
  in the on region and missing in the off region. The TRGB measured by
  \citet{Jackson+07b} and \citet{Boyer+09b} is marked with a dashed
  line, and the 75\% completeness limit is marked with a dotted
  line. The gray histogram shows the luminosity function for the
  entire field of view.\label{fig:on_lfunc}}
\end{figure}

Most of the DUSTiNGS galaxies have smaller angular sizes than the
field-of-view (Table~\ref{tab:targets}) and the recovered photometry
is therefore dominated by foreground and/or background sources. We
demonstrate this in Figure~\ref{fig:lfunc} for Cetus,
Aquarius, and Sag\,DIG. Cetus is far from the Galactic
Plane ($b = -73\degr$; Fig.~\ref{fig:targs}), so is dominated by red
background sources.  This causes the 4.5~\micron\ luminosity function
to appear brighter than the 3.6~\micron\ luminosity function. It also
results in a sharp drop-off near 17~mag, because most of the brighter
background galaxies have been eliminated from the GSC
(Section~\ref{sec:gal}). At the other extreme, Sag\,DIG is along a
line-of-sight near the Galactic Bulge ($b = -16\degr$). Foreground
therefore dominates its luminosity function and since these stars have
colors near zero, the luminosity function is nearly the same at both
wavelengths. Aquarius is at an intermediate latitude and shows the
signatures of both foreground and background sources.

In Figure~\ref{fig:on_lfunc}, we show the 3.6~\micron\ luminosity
function for on and off regions towards IC\,1613, which is known to
harbor a large intermediate-aged stellar population
\citep[e.g.,][]{Skillman+2014}. At $R < 4\arcmin$, the TRGB and a
feature attributed to x-AGB stars (Section~\ref{sec:xagb_class}) are
visible. These same features are visible in other
  galaxies with large AGB populations. For $R > 7\arcmin$, a feature
attributable to background sources is visible from $17 < m_{\rm [3.6]}
< 18$~mag.

\section{The IR Stellar Populations}
\label{sec:pops}

We cannot separate member stars from
  background/foreground sources with only the DUSTiNGS
  wavelengths. Therefore, we statistically subtract foreground and
  background sources to estimate the sizes of the TP-AGB
  ($N_{\rm TRGB}$) and x-AGB ($N_{\rm xAGB}$) populations. In
  Paper\,II, we use the 2-epoch variability information to identify a
  subset of individual AGB stars.

\subsection{Stellar Classification}
\subsubsection{AGB Stars ($N_{\rm TRGB}$)}
\label{sec:agb_class}

We classify all sources brighter than the TRGB as TP-AGB candidates, and
assume that the TRGB lies at $M_{3.6}=-6$~mag. The TRGB is unknown for
most of the DUSTiNGS galaxies, but \citet{Jackson+07a,Jackson+07b} and
\citet{Boyer+09b} find that it is $-6.6 < M_{3.6} < -6$~mag for 8 of
the DUSTiNGS dIrr galaxies. Using the Padova stellar evolution models
\citep{Marigo+2008,Marigo+13}, \citet{Bruzual+2013} created simple
stellar population models of the Magellanic Clouds.  They find that
$>$90\% of the thermally-pulsing (TP-)AGB stars are brighter than the
TRGB (G.~Bruzual 2013, private communication), so using the TRGB
cutoff ensures that most of TP-AGB stars are included here. We
apply no additional color cuts to the general TP-AGB classification.

The photometry is not 100\% complete at the assumed TRGB for most of
the DUSTiNGS galaxies. We therefore include a completeness-corrected
value of $N_{\rm TRGB}$ in Table~\ref{tab:stats} (see below). We do
not, however, correct for intrinsic crowding, which affects only the
inner 1\arcmin\ of IC\,10, NGC\,147, and NGC\,185.

The parameter $N_{\rm TRGB}$ includes AGB stars, massive young stars,
and massive evolved stars.  Without data at shorter wavelengths, it is
impossible to know what fraction of $N_{\rm TRGB}$ is indeed AGB
stars.  In the Magellanic Clouds, AGB stars account for 38\% (LMC) to
43\% (SMC) of the stars brighter than $-6$~mag \citep[derived from
  SAGE data after subtraction of foreground sources;][]{Boyer+11}.
For galaxies with recent star formation (i.e., the dIrr galaxies and
NGC\,185 and NGC\,147), we expect that the number of AGB candidates is
$\gtrsim$0.3$N_{\rm TRGB}$, based on the LMC and SMC
results. In the quiescent galaxies (i.e., most of the
  dSph galaxies), we can be confident that all, or nearly all, of
  $N_{\rm TRGB}$ are AGB candidates. Stars more massive than $M
  \gtrsim 8~M_\odot$ will not go through the AGB phase, so unless star
  formation has occurred in the last 50~Myr, there will not be
  contamination from massive stars in $N_{\rm TRGB}$.

\subsubsection{x-AGB Stars ($N_{\rm xAGB}$)}
\label{sec:xagb_class}

The x-AGB stars are a very dusty subset of the general
  TP-AGB population ($N_{\rm TRGB}$ includes $N_{\rm xAGB}$). More
  than 90\% of TP-AGB stars with $[3.6]-[4.5]>0.1$~mag and
  $M_{3.6}=-8$~mag in the Magellanic Clouds are classified as x-AGB
  stars by \citet{Blum+06} and \citet{Boyer+11}, and we use the same
  criteria to classify them here. We emphasize that the x-AGB label is not
  synonymous with dust-producing, nor is it exclusive; TP-AGB
  stars with bluer colors may be producing dust, though at a
  smaller rate \citep{Riebel+2012,Boyer+2012}. This x-AGB star
  classification is based solely on the observed IR color, and it
  roughly corresponds to AGB sources that are in the superwind phase,
  when the mass-loss rate exceeds the nuclear-consumption rate and the
  dust-production rate can increase by more than a factor of 10.

For galaxies observed with the longest total exposure
  times ($t_{\rm exp}=1080$~s), the magnitude uncertainties for x-AGB
  stars is $\lesssim$0.04~mag (1\,$\sigma$; Fig.~\ref{fig:errors}), so
  a color of $[3.6]-[4.5] = 0.1$~mag has a significance of
  $\gtrsim$2.5\,$\sigma$. Therefore, the x-AGB class will include some
  sources that are not truly dusty and vice versa. For galaxies with
  shorter total exposure times ($t_{\rm exp} = 60$~s and 150~s), the
  photometric uncertainties are larger and lie around 0.1~mag. In
  these cases, any infrared excess will have less
  significance. However, none of the galaxies with medium and shallow
  total exposure times show evidence for {\it any} sources redder than
  $[3.6]-[4.5] = 0.1$~mag, AGB or otherwise (Table~\ref{tab:stats}).

$N_{\rm xAGB}$ excludes most of the AGB stars with low mass-loss
rates, massive red supergiant stars, and massive main-sequence stars. The result
thus provides an estimate of the number of (mostly C-rich) x-AGB
stars, with limited contamination from other source types
\citep[cf.][]{Bonanos+2010,Boyer+11,Sewilo+2013}.

We caution that the notation used for dusty AGB stars varies. For
example, \citet{Gruendl+2008} reserve the term ``extreme AGB stars''
for the rarest, dustiest stars with $[3.6]-[4.5] \gtrsim 3$~mag.

\subsection{Background/Foreground Source Subtraction}
\label{sec:spat}

\begin{deluxetable*}{lrrrrlrrrr}
\tablewidth{\textwidth}
\tabletypesize{\normalsize}
\tablecolumns{10}
\tablecaption{AGB Population Size\label{tab:stats}}

\tablehead{
&
\multicolumn{2}{c}{--- Raw Counts ---}&
\multicolumn{2}{c}{--- Corrected ---}&
&
\multicolumn{2}{c}{--- Raw Counts ---}&
\multicolumn{2}{c}{--- Corrected ---}\\
\colhead{Galaxy}&
\colhead{$N_{\rm TRGB}$\tablenotemark{a}}&
\colhead{$N_{\rm xAGB}$\tablenotemark{b}}&
\colhead{$N_{\rm TRGB}$\tablenotemark{a}}&
\multicolumn{1}{c}{$N_{\rm xAGB}$\tablenotemark{b}}&
\colhead{Galaxy}&
\colhead{$N_{\rm TRGB}$\tablenotemark{a}}&
\colhead{$N_{\rm xAGB}$\tablenotemark{b}}&
\colhead{$N_{\rm TRGB}$\tablenotemark{a}}&
\colhead{$N_{\rm xAGB}$\tablenotemark{b}}
}

\startdata
And\,I       & $168\pm33$ & $\leq 7$  & $197\pm36$ & $\leq 8$  &     Coma         &   $\leq 3$&       $0$ &    $\leq 3$ & $0$ \\                   
And\,II      &  $73\pm31$ &  $9\pm4$  &  $86\pm34$ & $11\pm4$  &     CVn\,II      &   $\leq 5$&       $0$ &    $\leq 5$ & $0$ \\                   
And\,III     & $136\pm31$ & $\leq 6$  & $163\pm34$ & $\leq 7$  &     Hercules     &   $20\pm9$&       $0$ &    $24\pm10$ & $0$ \\                   
And\,V       &  $71\pm39$ & $\leq 8$  &  $77\pm43$ & $\leq 8$  &  IC\,10$^\dagger$ &$11\,200\pm137$& $516\pm23$ & $16\,996\pm158$ & $597\pm25$ \\
And\,VI      & $160\pm30$ & $\leq 6$  & $190\pm33$ & $\leq 6$  &     IC\,1613     &$2224\pm85$& $64\pm10$ & $2607\pm91$ & $67\pm11$ \\       
And\,VII     & $506\pm48$ & $\leq 10$ & $628\pm54$ & $\leq 11$ &     Leo\,A       &  $53\pm28$&  $\leq 5$ &   $63\pm31$ & $\leq 5$ \\              
And\,IX      & $\leq 44$  & $\leq 3$  &  $\leq 47$ & $\leq 3$  &     Leo\,IV      &   $\leq 6$&       $0$ &    $\leq 6$ & $0$ \\                   
And\,X       & $227\pm35$ & $\leq 6$  & $266\pm38$ & $\leq 6$  &     Leo\,T       &  $32\pm12$&  $\leq 1$ &   $36\pm13$ & $\leq 1$ \\              
And\,XI      &  $95\pm30$ & $\leq 7$  & $110\pm33$ & $\leq 7$  &     Leo\,V       &   $\leq 5$&       $0$ &    $\leq 6$ & $0$ \\                   
And\,XII     & $110\pm36$ & $\leq 9$  & $132\pm40$ & $\leq 9$  &     LGS\,3       &  $\leq 34$&  $\leq 3$ &   $\leq 36$ & $\leq 3$ \\             
And\,XIII    & $119\pm31$ & $\leq 6$  & $146\pm34$ & $\leq 6$  &NGC\,147$^\dagger$ &$4646\pm88$& $109\pm12$& $6342\pm100$& $124\pm13$ \\ 
And\,XIV     &  $50\pm29$ & $\leq 5$  &  $58\pm31$ & $\leq 5$  &NGC\,185$^\dagger$ &$4119\pm78$& $86\pm10$ & $5180\pm86$ & $99\pm11$ \\  
And\,XV      &  $46\pm26$ & $\leq 4$  &  $55\pm29$ & $\leq 5$  &     Pegasus      & $742\pm54$& $\leq 11$ &  $882\pm58$ & $\leq 12$ \\          
And\,XVI     &  $40\pm30$ & $\leq 1$  &  $46\pm21$ & $\leq 1$  &     Phoenix      &  $61\pm16$&  $\leq 2$ &   $68\pm17$ & $\leq 3$ \\    
And\,XVII    & $128\pm38$ & $\leq 9$  & $150\pm41$ & $\leq 10$ &     Pisces\,II   & $\leq 9$  &       $0$ &    $\leq 10$ & $0$ \\ 
And\,XVIII   & $317\pm53$ & $\leq 24$ & $406\pm60$ & $\leq 26$ &     Sag\,DIG     & $829\pm79$& $\leq 26$ & $1239\pm92$ & $\leq 29$ \\ 
And\,XIX     & $\leq 62$  & $\leq 9$  &  $\leq 67$ &  $\leq 9$ &     Segue\,1     &   $\leq 1$&       $0$ &    $\leq 1$ & $0$ \\                   
And\,XX      & $130\pm30$ & $\leq 7$  & $157\pm33$ &  $\leq 8$ &     Segue\,2     &   $\leq 2$&       $0$ &    $\leq 3$ & $0$ \\                   
And\,XXI     & $116\pm39$ & $\leq 10$ & $135\pm43$ &  $\leq 11$&     Segue\,3     &   $\leq 1$&       $0$ &    $\leq 2$ & $0$ \\                   
And\,XXII    &  $99\pm36$ & $\leq 8$  & $122\pm40$ &  $\leq 9$ &     Sextans\,A   & $965\pm79$& $\leq 34$ & $1230\pm88$ & $\leq 37$ \\   
Antlia       & $204\pm48$ & $\leq 23$ & $260\pm54$ & $\leq 25$ &     Sextans\,B   &$1613\pm75$& $77\pm20$ & $2118\pm86$ & $88\pm22$ \\
Aquarius     & $205\pm75$ & $\leq 14$ & $253\pm53$ & $\leq 15$ &     Tucana       & $150\pm35$&  $\leq 6$ &  $183\pm38$ & $\leq 6$ \\ 
Bootes\,I    & $\leq 8$   & $0$       &  $\leq 8$  &       $0$ &     UMa\,II      &   $\leq 2$&       $0$ &    $\leq 2$ & $0$ \\                   
Bootes\,II   &      $0$   & $0$       &       $0$  &       $0$ &     Willman\,1   &        $0$&       $0$ &         $0$ & $0$ \\                   
Cetus        & $140\pm29$ & $7\pm4$   & $166\pm31$ &   $9\pm4$ &     WLM          &$1764\pm72$& $59\pm12$ & $2077\pm78$ & $67\pm13$

\enddata 

\tablenotetext{a}{\ Stars that are brighter than $M_{\rm 3.6} =
  -6$~mag. Depending on the star-formation history of the galaxy, the
  total number of AGB stars can range from $0.3\,N_{\rm TRGB}$ --
  $N_{\rm TRGB}$ (see text).}
  
\tablenotetext{b}{\ xAGB stars are those brighter than $M_{\rm 3.6} =
  -8$~mag and redder than $[3.6]-[4.5]=0.1$~mag.}

\tablenotetext{$\dagger$}{\ These galaxies are affected by
    intrinsic crowding in their centers (Table~\ref{tab:crowd}), so $N_{\rm
      TRGB}$ should be considered a lower limit in these
    cases. Crowding does not affect $N_{\rm xAGB}$ except within the
    central $\approx$1\arcmin\ region of IC\,10. We have not corrected
    numbers in this table for intrinsic crowding.}

\tablecomments{\ The size of the stellar population derived by
  subtracting the background and foreground contamination. Upper
  limits at 95\% confidence are quoted when AGB stars are not detected
  above the level of background$+$foreground sources. The sources
  included here are confined to the spatial area covered by all epochs
  and wavelengths (Table~\ref{tab:obs}). We report both the raw
    counts and the counts corrected for photometric completeness
    (Section~\ref{sec:comp}).}
\end{deluxetable*}

Each DUSTiNGS galaxy was observed with a large field of view to
assist in subtracting the contribution of background and foreground
sources.  To estimate $N_{\rm TRGB}$ and $N_{\rm xAGB}$, we first
determine the distance from each galaxy center where the radial
profile of point sources becomes flat and measure the density of
sources with the relevant colors and magnitudes beyond this distance
($\Sigma_{\rm N}$).  We then subtract $\Sigma_{\rm N} \times {\rm
  coverage\ area}$ (Table~\ref{tab:obs}) from the total number of
point sources to obtain $N_{\rm TRGB}$ and $N_{\rm xAGB}$.

In regions where the stellar density is high, background galaxies are
undetectable. In these regions, we subtract only the foreground
sources, which we estimate for the position of each target galaxy
using the TRILEGAL population synthesis code \citep[][see
  Fig.~\ref{fig:fg}]{Girardi+05}. Table~\ref{tab:stats} lists the
resulting AGB population sizes. The uncertainties in these numbers are
derived from background-limited Poisson statistics.  If the number of
sources is below the 1.6$\sigma$ limit, we quote 95\% confidence upper
limits. 

Table~\ref{tab:stats} includes both the raw values of
  $N_{\rm TRGB}$ and $N_{\rm xAGB}$ and values that have been
  corrected for photometric completeness using each galaxy's
  completeness curve (Fig.~\ref{fig:complim} shows the mean
  completeness curve for each photometric depth). To make this
  correction, we first apply the completeness curve to the total
  number of counts and to $\Sigma_{\rm N}$ individually, then compute
  $N_{\rm TRGB}$ and $N_{\rm xAGB}$ from those corrected values.

\subsection{Dust Production at Very Low Metallicity}

While most of the x-AGB stars in the DUSTiNGS sample are in the
massive, more metal-rich galaxies (IC\,10, NGC\,147, NGC\,185, and
WLM), we find 166$\pm$28 x-AGB stars at ${\rm [Fe/H]} \approx -1.6$
and $9\pm4$ at ${\rm [Fe/H]} \approx -1.9$ (Cetus). These are some of
the most metal-poor dusty AGB stars known, and they are likely to be
C-rich. AGB stars in the SMC with similar $[3.6]-[4.5]$ colors have an
average dust-production rate of $\log(\dot{D})
= -8.7\ [M_\odot/{\rm yr}]$ \citep{Boyer+2012}.

For galaxies with ${\rm [Fe/H]} < -2$, we can quote only upper limits
for the number of x-AGB stars.  On the other hand, we do detect
1645$\pm$240 AGB stars with less dust in these metal-poor galaxies
(And\,XI, And\,XII, And\,XIV, And\,XVI, Hercules, Leo\,T,
Sag\,DIG). In the SMC, AGB stars at these colors have dust-production
rates of $-10.7 < \log(\dot{D}) < -10.1\ [M_\odot/{\rm yr}]$.
Because the x-AGB population sizes are detected
  statistically, we can say little about the properties of the
  individual stars (e.g., their distribution in color and luminosity
  and their dust-production rates).  In Paper\,II, we identify a
  subset of the individual x-AGB stars and further describe their
  characteristics.

\section{Conclusions}
\label{sec:concl}

DUSTiNGS is a 3.6 and 4.5~\micron\ photometric survey of 50 resolved
dwarf galaxies within 1.5~Mpc designed to detect dusty evolved stars.
The survey includes 37 dSph galaxies, 8 dIrr galaxies, and 5 dIrr/dSph
transition-type galaxies.  The large sample size allows for robust
statistics on the short-lived, dust-producing phase. Each galaxy was
observed over two epochs to aid in identifying variable AGB
stars; Paper\,II presents the results of the variability
analysis. Here, we describe the targets, the observing strategy, and
the publicly-available data products.

For all galaxies, the photometry is $>$75\% complete within the
possible magnitude range of the TRGB with the exception of the inner
regions of the most crowded galaxies: IC\,10, NGC\,147, and
NGC\,185. This completeness enables the detection of most of the AGB
and massive evolved star populations. The photometric catalogs are
publicly available at MAST, VizieR, and IRSA.

Because it is difficult to distinguish dusty evolved stars from
unresolved background objects at these wavelengths, the DUSTiNGS
survey imaged an area larger than the half-light radius of each
galaxy to allow for statistical subtraction of foreground and
background sources. We present here an estimate of the size of the
stellar population brighter than the TRGB and the size of the dusty
AGB star population. We find $1062\pm103$ ``extreme'' dusty
AGB stars in 21 of the DUSTiNGS galaxies. For the remaining 29
DUSTiNGS galaxies we report 95\% confidence upper limits.

\acknowledgements

Many thanks to Brian Babler for very helpful discussions about IRAC
photometry.  We also thank the referee for his/her helpful
comments. This work is supported by {\it Spitzer} via grant GO80063
and by the NASA Astrophysics Data Analysis Program grant number
N3-ADAP13-0058. MLB is supported by the NASA Postdoctoral Program at
the Goddard Space Flight Center, administered by ORAU through a
contract with NASA. RDG was supported by NASA and the United States
Air Force. AZB acknowledges funding by the European Union (European
Social Fund) and National Resources under the ``ARISTEIA'' action of
the Operational Programme ``Education and Lifelong Learning'' in
Greece. GCS receives support from the NSF, award AST-1108645.


\appendix
\section{Color-magnitude Diagrams}

Figures~\ref{Afig:cmds1} and \ref{Afig:cmds2} show the DUSTiNGS
color-magnitude diagrams. We show the combined epochs to demonstrate
the maximum photometric depth. The dark shaded regions mark the range
of the expected TRGB for each galaxy. The majority of TP-AGB stars are
brighter than this limit. The light shaded regions mark the
approximate location of x-AGB stars.

\begin{figure*}
\includegraphics[width=\textwidth]{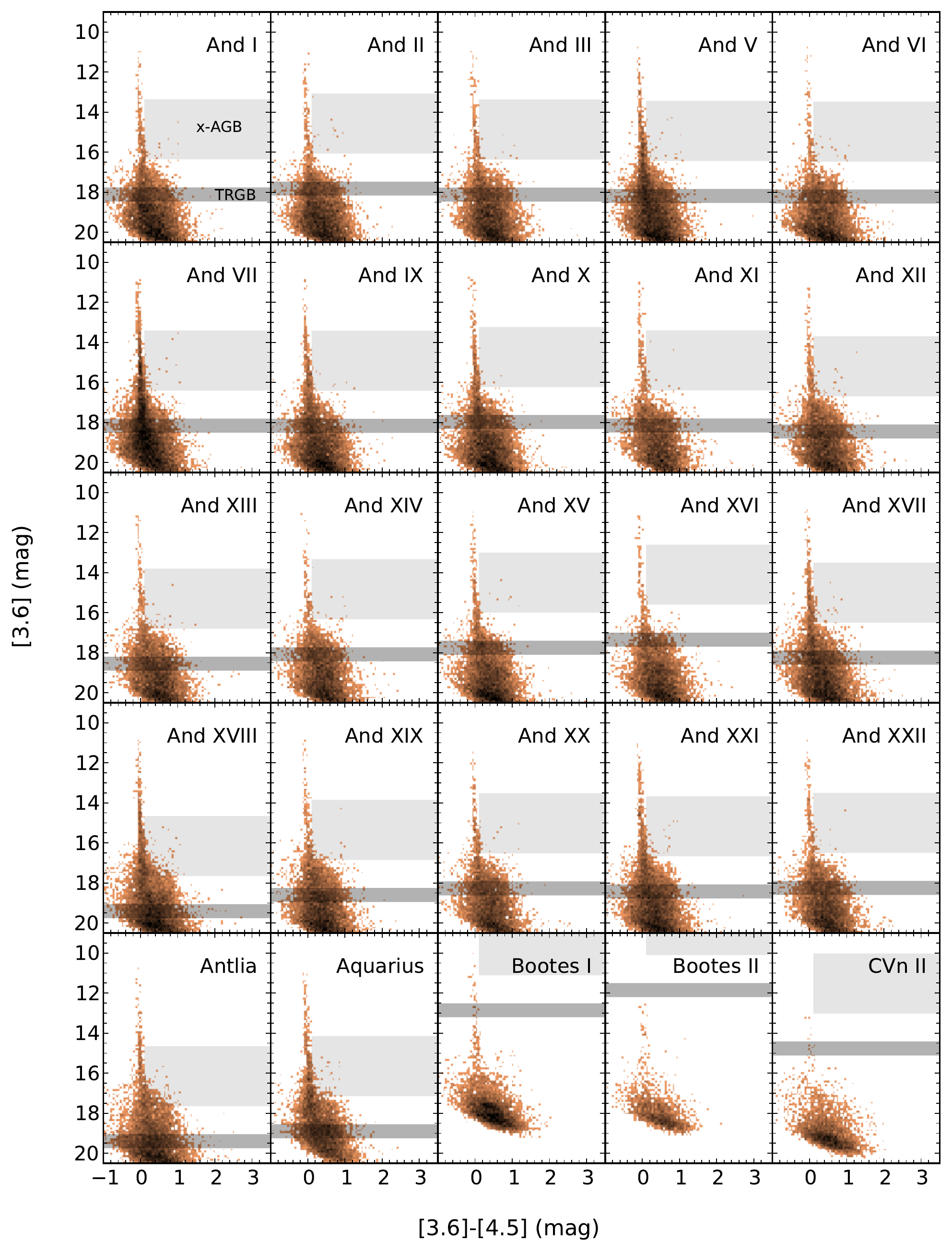}
\caption{Color-magnitude diagrams of the GSC for each DUSTiNGS galaxy.
  Magnitudes shown here are derived from the two combined epochs. The
  dark shaded region marks the range of the possible TRGB and the
  light shaded region marks the approximate location of x-AGB stars.
\label{Afig:cmds1}
}
\end{figure*}

\begin{figure*}
\includegraphics[width=\textwidth]{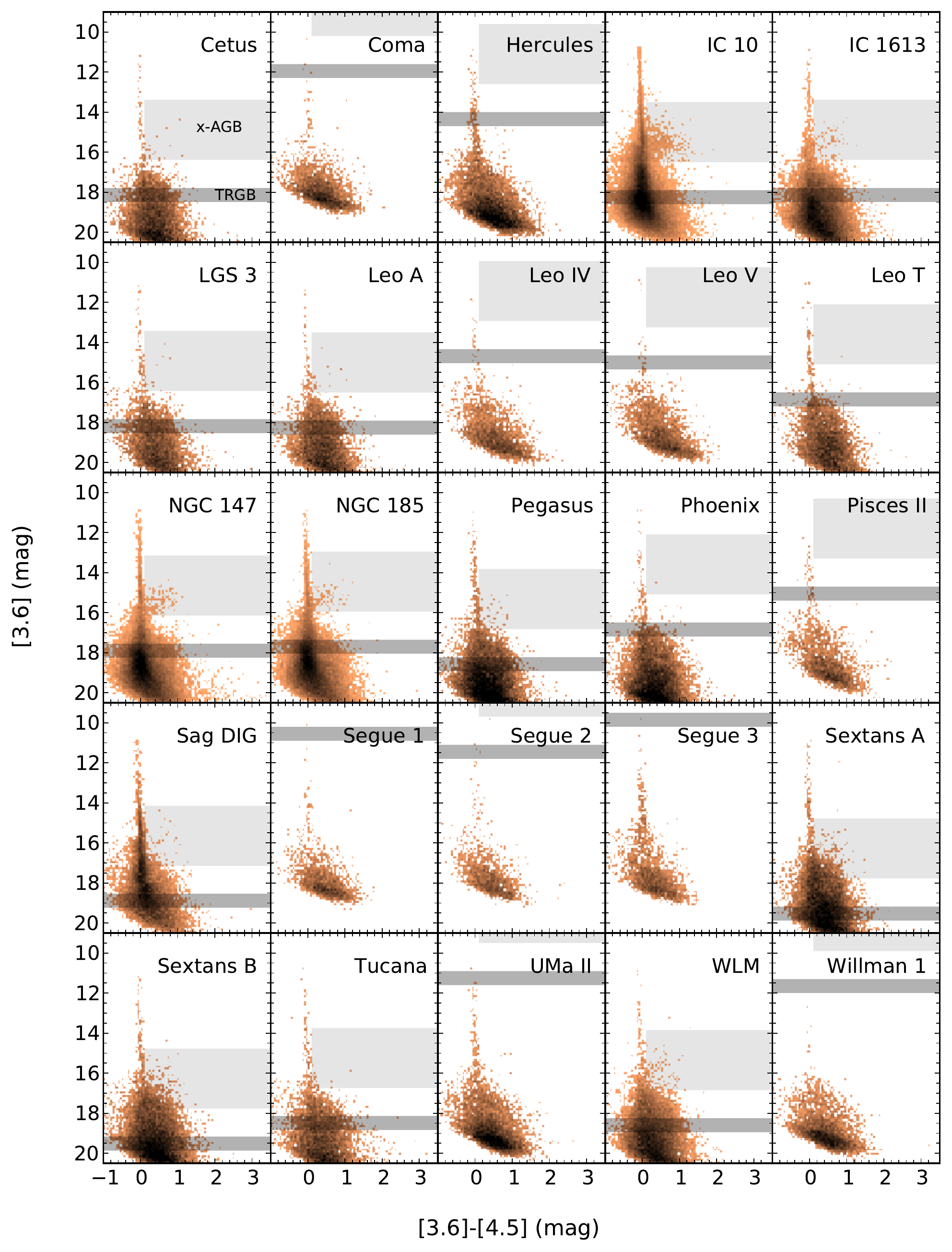}
\caption{Continued.
\label{Afig:cmds2}
}
\end{figure*}

\section{Images}

\begin{figure*}
  \includegraphics[width=\columnwidth]{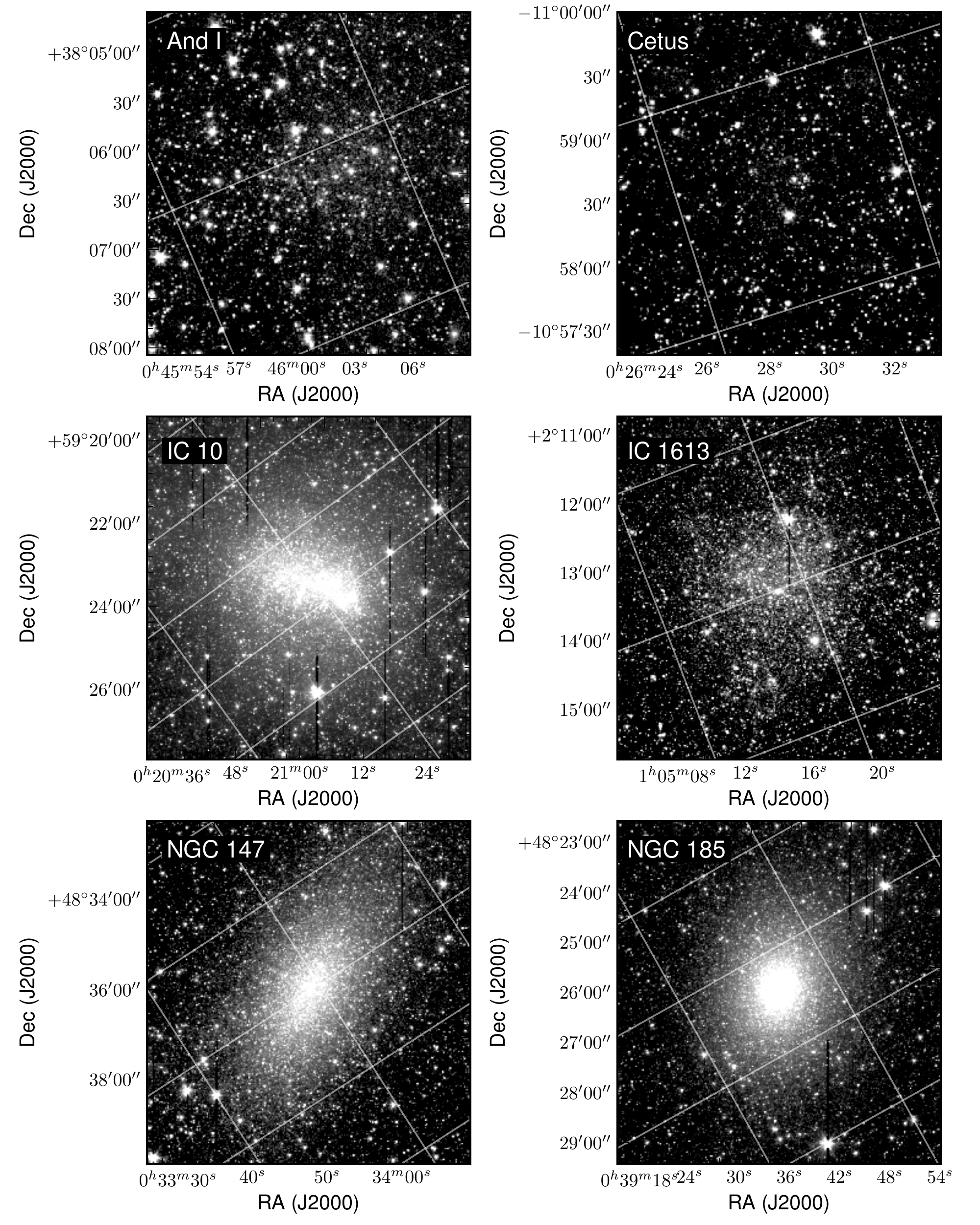}
\caption{3.6~\micron\ epoch~1 mosaics for a subset of the DUSTiNGS
  galaxies. \label{Afig:mos1}}
\end{figure*}

\begin{figure*}
  \includegraphics[width=\columnwidth]{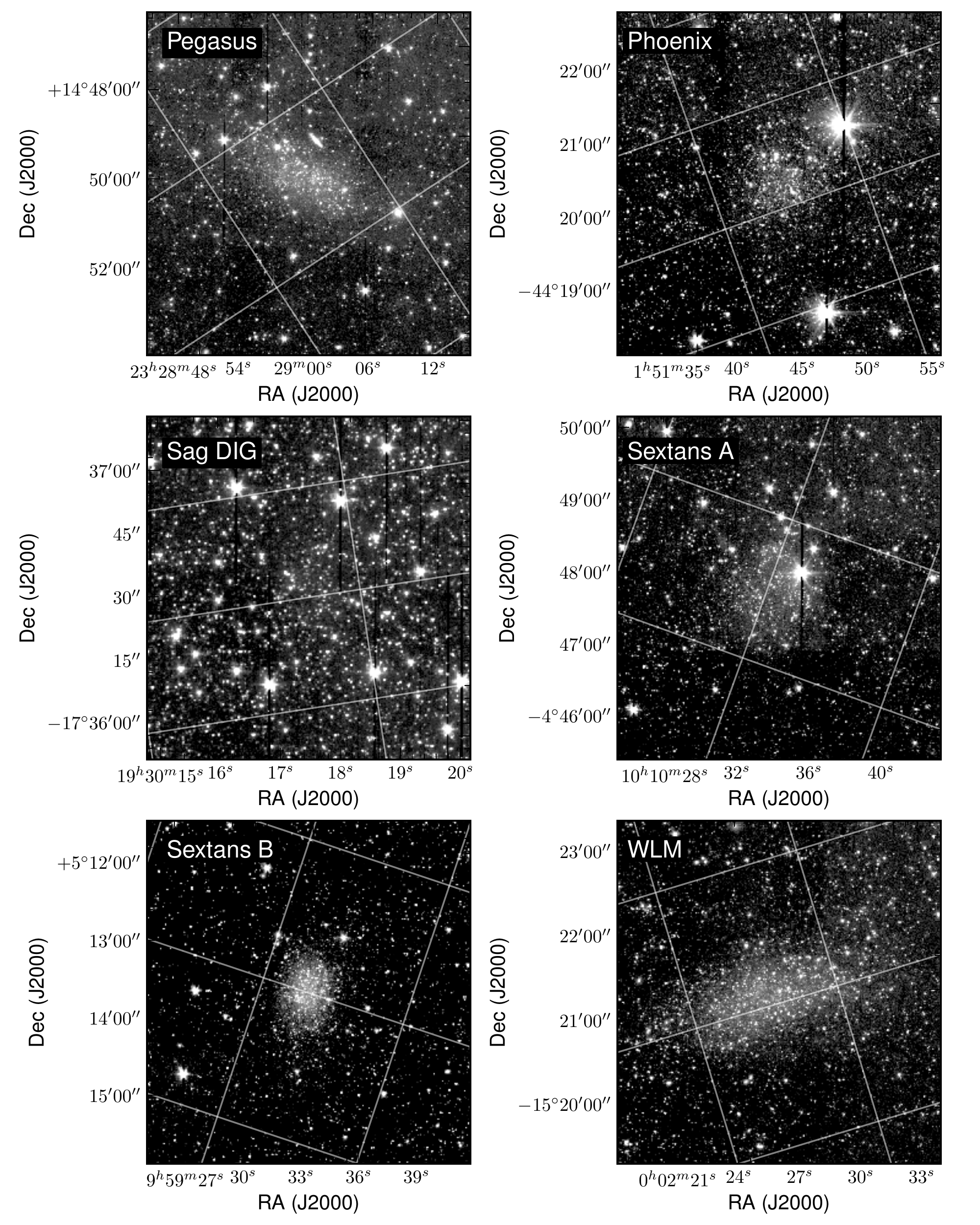}
\caption{Figure~\ref{Afig:mos1} continued. \label{Afig:mos2}}
\end{figure*}

Figures~\ref{Afig:mos1} and \ref{Afig:mos2} show the
3.6~\micron\ epoch~1 mosaics for a subset of the DUSTiNGS galaxies.
Galaxies not shown are low mass and have few sources above the
TRGB. For these galaxies, it is difficult to see the galaxy among the
background and foreground sources.  We include Cetus as an example of
a low-mass galaxy. See Figure~\ref{fig:img} for an example of the
imaging strategy.

\end{document}